\documentclass[reprint,twocolumn,ap,prb,nofootinbib,showpacs]{revtex4-1}
\usepackage{graphicx}
\usepackage{times}
\usepackage{bm}
\usepackage{natbib}
\usepackage{amsfonts}
\usepackage[linktocpage=true]{hyperref}

\newcommand*\xbar[1]{%
  \hbox{%
    \vbox{%
      \hrule height 0.5pt 
      \kern0.5ex
      \hbox{%
        \kern-0.1em
        \ensuremath{#1}%
        \kern-0.1em
      }%
    }%
  }%
} 

\begin{document}

\title{Robustness of Topological Superconductivity in Proximity-Coupled Topological Insulator Nanoribbons}
\author{ Piyapong Sitthison}
\affiliation{Department of Physics and Astronomy, West Virginia University, Morgantown, WV 26506, USA}
\author{Tudor D. Stanescu}
\affiliation{Department of Physics and Astronomy, West Virginia University, Morgantown, WV 26506, USA}

\begin{abstract}
We study the low-energy physics of topological insulator (TI) nanoribbons proximity-coupled to $s$-wave superconductors (SCs) by explicitly incorporating the proximity effects that emerge at the TI-SC interface. We construct a low-energy effective theory that incorporates the proximity effect through an interface contribution containing both normal and anomalous terms and an energy-renormalization matrix. We show that the strength of the proximity-induced gap is determined by the transparency of the interface and the amplitude of the low-energy TI states at the interface. Consequently, the induced gap is strongly band-dependent and collapses for bands containing states with low amplitude at the interface. We find that states with energies within the bulk TI gap have surface-type character and, in the presence of proximity-induced or applied bias potentials, have most of their weight near either the top or the bottom surface of the nanoribbon. As a result, single interface TI-SC structures are susceptible to experiencing a collapse of the induced gap whenever the chemical potential is far enough from the value corresponding to the bulk TI Dirac point and crosses weakly coupled bands.  We also find that changing the chemical potential in single-interface structures using gate potentials may be ineffective, as it does not result in a significant increase of the induced gap. On the other hand, we find that symmetric structures, such as a TI nanowire sandwiched between two superconductors, are capable of realizing the full potential of TI-based structures to harbor robust topological superconducting phases.
\end{abstract}


\maketitle

\section{Introduction}

The discovery of topological phases of matter -- phases having the same symmetry as the Hamiltonian and being characterized by certain quantities that remain invariant under small adiabatic deformations of the Hamiltonian -- promises to open a new major theme in condensed-matter physics. The interest in pursuing this exciting possibility was initiated by a series of theoretical advances\cite{Thouless1982,Haldane1998,Wen1995,Kane2005,Bernevig2006,Fu2007,Moore2007,Qi2008,Zhang2009} following the discovery of the quantum Hall effect (QHE)\cite{Klitzing1980,Tsui1982}, and it was galvanized in recent years by a number of experimental breakthroughs, such as the realization of the quantum spin Hall effect in Hg quantum wells\cite{Konig2007} and the discovery of three-dimensional (3D) topological insulators (TIs).\cite{Hsieh2008,Xia2009} The ultimate success of this new field will depend on our ability to exploit the properties of topological quantum systems for novel technological applications, e.g.,  using non-Abelian anyons for fault-tolerant topological quantum computation.\cite{Freedman1998,Kitaev2003,DasSarma2005,Nayak2008}  A very promising direction within the field of topological states of matter involves the realization and detection of topological superconducting phases with non-Abelian zero-energy Majorana bound states\cite{Majorana1937,Moore1991,Read2000,Kitaev2001,Wilczek2009} in hybrid solid-state structures. From a theoretical standpoint, the existence of zero-energy Majorana bound states localized at defects or near the boundaries in systems that support topological superconducting phases stands on firm ground.\cite{Alicea2012,Leijnse2012,Beenakker2013,Stanescu2013} There are many concrete proposals for realizing the conditions necessary for the emergence of Majorana bound states using proximity-coupled heterostructures, such as topological insulator - superconductor\cite{Fu2008,Fu2009,Tanaka2009,Linder2010,Cook2011,Cook2012} and semiconductor - superconductor\cite{Sau2010,Alicea2010,Sau2010a,Lutchyn2010,Oreg2010} hybrid structures.  On the experimental front, encouraging results were recently reported\cite{Mourik2012,Deng2012,Das2012,Rokhinson2012,Churchill2013,Finck2013,Lee2014} on the observation of signatures consistent with the presence of zero-energy Majorana modes in semiconductor nanowire - superconductor hybrid structures. Further progress requires the optimization of the conditions necessary for the emergence of topological superconductivity, to ensure the unambiguous demonstration of Majorana bound states and to enhance the robustness of the topological quantum phase,  which could eventually allow the controlled manipulation of the Majorana quasiparticles.\cite{Alicea2011,Clarke2011,vanHeck2012,Halperin2013}  

The key ingredients of the proposed semiconductor - superconductor (SM-SC) structure\cite{Sau2010,Alicea2010,Sau2010a,Lutchyn2010,Oreg2010} that has recently attracted a lot of attention\cite{Mourik2012,Deng2012,Das2012,Rokhinson2012,Churchill2013,Finck2013,Lee2014} are strong spin-orbit coupling, proximity-induced superconductivity, and Zeeman splitting. The stability of the zero-energy Majorana modes can be enhanced by increasing the quasiparticle gap that characterizes the topological superconducting state, which can be realized by increasing the strength of the spin-orbit coupling and the magnitude of the proximity-induced pair potential.\cite{Sau2010a,Potter2011,*Potter2011a} Building upon these ideas and taking advantage of the remarkably strong spin-orbit coupling that characterizes the recently discovered 3D topological insulators,\cite{Hsieh2008,Xia2009} Cook and Franz have proposed the realization of topological superconductivity and Majorana bound states using TI nanowires proximity-coupled to ordinary s-wave superconductors.\cite{Cook2011} In addition to the potential benefit of strong spin-orbit coupling, this proposal also addresses a  major challenge facing the practical implementation of the SM-SC idea: controlling the position of the chemical potential, which is required to lie in narrow windows near the bottoms of the confinement-induced SM bands.\cite{Lutchyn2010,Oreg2010,Lutchyn2011} More specifically, it was shown\cite{Cook2011,Cook2012} that zero energy Majorana bound states could persist for any value of the chemical potential inside the TI bulk band gap and that the Majorana modes are robust against strong non-magnetic disorder. The feasibility of this proposal is further supported by some recent experimental progress, which includes the synthesis of TI nanowires and nanoribbons\cite{Kong2010} and the observation of the superconducting proximity effect in Bi$_2$Se$_3$ bulk systems\cite{Yang2012} and nanoribbons.\cite{Zhang2011} 

The role of the TI nanowire band structure, which incorporates the effect of strong spin-orbit coupling, in the realization of robust Majorana states in the presence of a longitudinal magnetic field has been previously investigated.\cite{Cook2011,Cook2012} However, the implications of the other critical element -- the proximity effect at the TI-SC interface\cite{Stanescu2010,Black-Schaffer2013} -- were not yet addressed in any detail. Instead, a simple s-wave pairing potential was assumed to exist in the TI nanowire as a result of this proximity effect. Previous studies of semiconductor-based Majorana nanostructures have demonstrated that a more careful treatment of the proximity effects can reveal nontrivial low-energy properties, such as  the suppression of the induced superconducting gap due to proximity-induced interband coupling\cite{Stanescu2013a} and the band-selective coupling to the metallic lead that may be responsible for the observed soft gap feature.\cite{Stanescu2013x} Moreover, the SC proximity effect is strongly-dependent on the details of the interface and on the nature of the low-energy bands in the nanowire\cite{Stanescu2013}, e.g., the s-wave character of electron-doped semiconductor bands and the predominantly p-wave character of the low-energy TI states.\cite{Zhang2009} Finally, electrostatic effects, including interface-induced potentials,\cite{Zhang2010} can significantly affect the spatial distribution of the low-energy states, particularly in systems with no bulk-type carriers\cite{Galanakis2012} and, implicitly, modify the effective SC-TI coupling and the strength of the SC proximity effect. All these elements may affect the magnitude of the induced SC pairing potential and, ultimately, the robustness of the topological superconducting phase that hosts the zero-energy Majorana bound states. 

In this work we systematically study the proximity effect at interfaces between TI nanoribbons and s-wave superconductors, and we investigate its impact on the stability of the emerging topological superconducting phase.  By explicitly incorporating relevant elements that control this proximity effect, such as the p-wave character of the low-energy TI states affected by the coupling to the superconductor, interface- or gate-induced potentials, and details of the device architecture, we identify a number of potential challenges facing the practical implementation of the proximity-coupled TI nanowire proposal for realizing Majorana fermions, as well as possible solutions for overcoming these challenges. We find that single-interface TI-SC structures are not ideal for harboring robust topological superconducting phases because of the presence of weakly coupled bands characterized by states that have most of their weight near surfaces other than the TI-SC interface. Consequently, the induced quasiparticle gap collapses whenever the chemical potential crosses such a weakly coupled band. On the other hand, we find that two-interface structures,  such as a TI nanowire sandwiched between two superconductors, do not generate this type of problem and  are capable of hosting exceptionally robust topological superconducting phases. 

The paper is organized as follows. In Section \ref{SecII} we describe the derivation of the low-energy effective theory used in this work, including the tight-binding model for the TI nanoribbon (Section \ref{SecIIA}) and the effective model that incorporates the proximity effect induced by the bulk SC (Section \ref{SecIIC}). The results of our numerical study of the low-energy model are reported in Section \ref{SecIII}. First, in Section \ref{SecIIIA}, we discuss some generic properties of the low-energy spectrum  in both normal and superconducting phases. We also investigate the real-space structure of the low-energy states and the effect of applying a bias potential on their spatial profile. In Section \ref{SecIIIB} we calculate the topological phase diagram and discuss the dependence of the phase boundaries on various relevant parameters. The dependence of the proximity-induced gap on the relevant control parameters for different TI-SC structures as well as the implications of these findings concerning the robustness of the topological phase are discussed in Section \ref{SecIIIC}. Our conclusions are presented in Section \ref{SecIV}. 

\section{Modeling of TI -- superconductor hybrid structures} \label{SecII} 

\subsection{Tight-binding model for TI nanoribbons} \label{SecIIA} 

\begin{figure}[tbp]
\begin{center}
\includegraphics[width=0.48\textwidth]{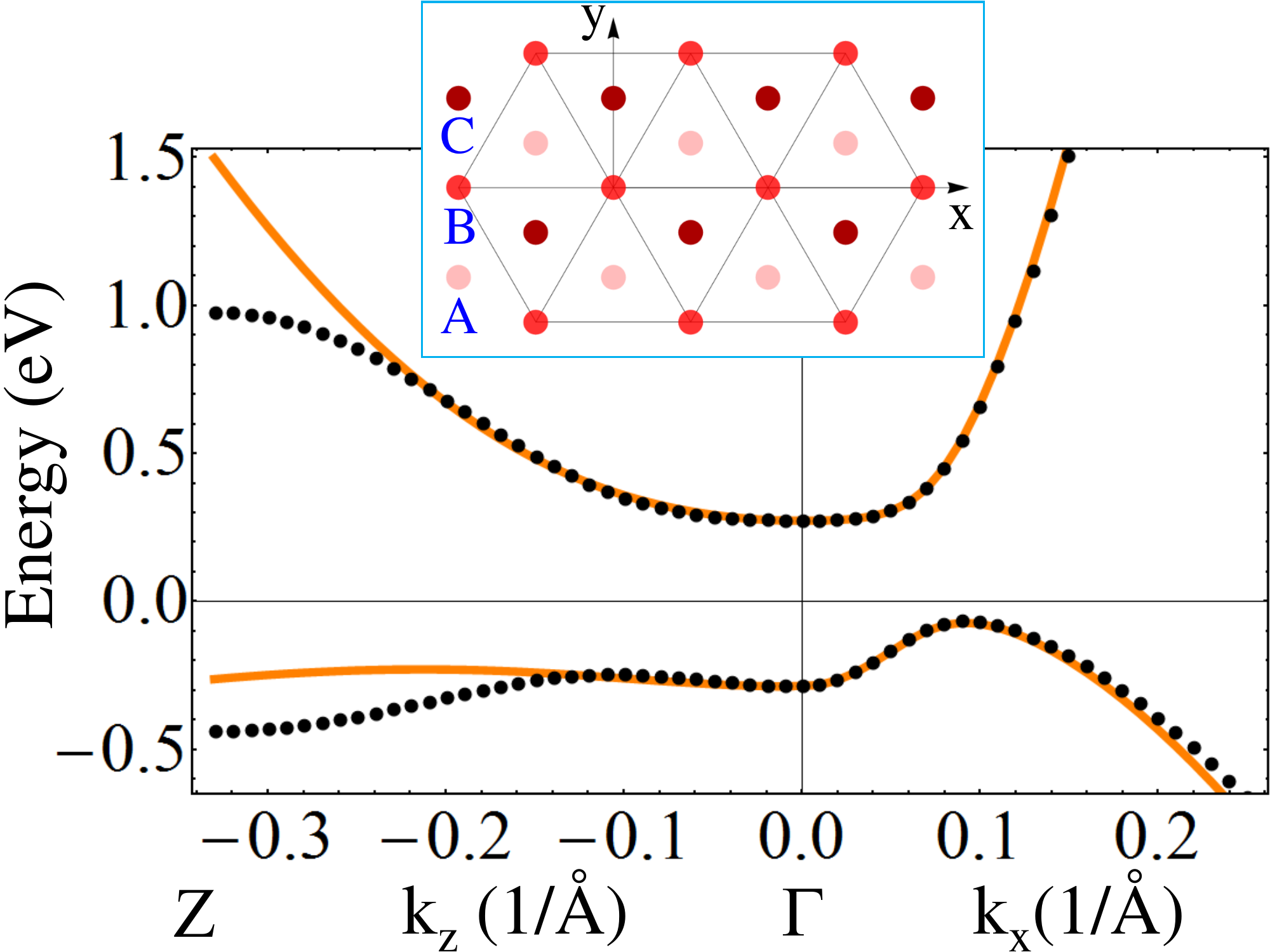}
\vspace{-5mm}
\end{center}
\caption{(Color online) Comparison between the low-energy spectrum of the 3D continuum model given by Eq. (\ref{HTI}) (orange/light gray lines) and that of the tight-binding model (\ref{HTIlatt}) (black dots).  The parameters of the lattice model were optimized, starting from the values given by Eq. (\ref{tij}), to capture the correct dispersion along the $\Gamma - Z$ direction in the Brillouin zone, which controls the inter-subband spacing in the spectrum of a TI thin film (see Fig. \ref{Fig2}). The relevant parameters, corresponding to Bi$_2$Se$_3$,  are: ($\epsilon_+$,  $\epsilon_-$, $t_{1+}$,  $t_{1-}$, $t_{2+}$,  $t_{2-}$, $\alpha_{1}$,  $\alpha_{2}$) $=$ ($17.6$, $-3.26$, $-0.11$, $0.06$, $-2.88$, $0.53$, $0.04$, $0.26$). The inset shows the structure of the rhombohedral lattice  (of lattice constants $a=4.138$\AA ~and $c=28.64$\AA) used in the construction of  the tight-binding model. The triangular sub-lattices correspond to Bi$_2$Se$_3$ quintuple layers and are staked in the $z$-direction.}
\vspace{-3mm}
\label{Fig1}
\end{figure}

To construct a low-energy effective theory for topological insulator (TI) nanowires proximity coupled to s-wave superconductors (SCs),  we start with a minimal tight-binding model that captures the key low-energy band-structure properties of 3D TIs from the Bi$_2$Se$_3$ family.\cite{Qi2011} According to first-principle calculations,\cite{Zhang2009,Zhang2010a}  the Bi$_2$Se$_3$ spectrum is characterized by a gap of about $0.3$eV separating the conduction and valence bands at the $\Gamma$ point (i.e. at ${\bm k}=0$).  In the vicinity of ${\bm k}=0$ the low-energy states are, predominantly, superpositions of Bi and Se p-type orbitals.\cite{Zhang2009,Liu2010} More specifically, the states near the bottom (top) of the conduction (valence) band have the form\cite{Liu2010}
\begin{eqnarray}
|\lambda, \uparrow\rangle &=& u_\lambda |\lambda, p_z, \uparrow\rangle + v_\lambda |\lambda, p_+, \downarrow\rangle, \nonumber \\
|\lambda, \downarrow\rangle &=& u_\lambda^* |\lambda, p_z, \downarrow\rangle + v_\lambda^* |\lambda, p_-, \uparrow\rangle, \label{states}
\end{eqnarray} 
where $\lambda = \pm 1$, $p_\pm=p_x \pm p_y$, and $u_\lambda$,  $v_\lambda$ (with $|u_\lambda|^2+|v_\lambda|^2=1$) are certain coefficients that depend on the spin-orbit coupling strength.\cite{Liu2010} In this work, we will not consider specific values for these coefficients, but instead we will treat them as variable model parameters. The states $|\lambda, \sigma\rangle$ in Eq. \ref{states} have parity $\lambda$ and are eigenstates of the total angular momentum along the $z$ direction with eigenvalues $\sigma\hbar/2$. The molecular orbitals $|\lambda, p_\alpha, \sigma\rangle$ have well-defined parity ($\lambda$) and spin ($\sigma$, in units of $\hbar/2$) and represent symmetric/antisymmetric superpositions of Bi and Se $p_\alpha$ orbitals that extend across a quintuple layer. One quintuple layer can be modeled by defining a tight-binding model on a triangular lattice with a basis given by the four states in Eq. \ref{states}. The full 3D model is defined on the rhombohedral lattice obtained by staking the triangular sublattices in the $z$ direction with three distinct positions of the sublattices, which results in an A-B-C type pattern\cite{Zhang2009,Zhang2010a,Liu2010} (see the inset of Fig. \ref{Fig1}). In the ${\bm k}\rightarrow 0$ limit, the low-energy spectrum is described by the continuum k$\cdot$p Hamiltonian\cite{Zhang2009}
\begin{equation}
{\cal H}_{\rm TI}({\bm k}) = \epsilon({\bm k}) + {\cal M}({\bm k}) \lambda_z + A_1 k_z \lambda_x\sigma_z +A_2\lambda_x(k_x\sigma_x+k_y\sigma_y), \label{HTI}
\end{equation}
where $\lambda_i$ and $\sigma_i$ are Pauli matrices associated with the orbital and spin degrees of freedom, respectively, $\epsilon({\bm k}) = C_0+C_1k_z^2+C_2k_\parallel^2$,  and ${\cal M}({\bm k})=M_0+M_1 k_z^2+M2_2k_\parallel^2$, with $k_\parallel^2=k_x^2+k_y^2$. The parameters of the effective model are obtained by fitting the low-energy spectrum obtained from {\it ab initio} calculations with the spectrum of the effective Hamiltonian.\cite{Zhang2009,Liu2010} The corresponding lattice model, which is written in the basis $\{|+,\uparrow\rangle$, $|-,\uparrow\rangle$, $|+,\downarrow\rangle,  |-,\downarrow\rangle\}$,  is given by the tight-binding Hamiltonian\cite{Hutasoit2011} 
\begin{equation}
H_{\rm TI} = \sum_{\lambda, i, j}\left[\left(\epsilon_{\lambda} \delta_{ij}+t_{ij}^{(\lambda)}\right)c_{i\lambda}^{\dagger}c_{j\lambda} -i \alpha_{ij} c_{i\lambda}^{\dagger}(\vec{\bm \delta}_{ij}\cdot\vec{\bm \sigma})c_{j\bar{\lambda}}\right], \label{HTIlatt}
\end{equation}
where $i$ and $j$ are sites on a rhombohedral lattice with lattice constants $a$ (distance between nearest-neighbor sites in a triangular sublattice)  and $c$ (with $c/3$ being the distance between two adjacent sublattices), $c_{i\lambda}^{\dagger}=(c_{i\lambda\uparrow}^{\dagger}, c_{i\lambda\downarrow}^{\dagger})$ are creation operators for states given by Eq. (\ref{states}) localized near site $i$, $\bar{\lambda}=-\lambda$, and $\vec{\bm \sigma} = (\sigma_x, \sigma_y, \sigma_z)$  is the Pauli vector. In Eq. (\ref{HTIlatt}) the ``hopping vector'' $\vec{\bm \delta}_{ij}=({\bm r}_j-{\bm r}_i)/a$  takes six values that correspond to the nearest-neighbors in the $x-y$ plane (see the inset of Fig. \ref{Fig1}), i.e.  $(\pm 1, 0, 0)$ and $(\pm 1/2, s \sqrt{3}/2, 0)$, with $s=\pm 1$, and six values  that correspond to out-of-plane hoppings, i.e. $(0, s/\sqrt{3}, s c/3a)$ and $(\pm 1/2, -s/2\sqrt{3}, s c/3a)$. The hopping parameters $t_{ij}^{(\lambda)}=t_{1\lambda}$ and $t_{ij}^{(\lambda)}=t_{2\lambda}$ correspond to nearest-neighbor out-of-plane and in-plane hopping, respectively, between $\lambda$-type molecular orbitals. The second term, representing the spin- and direction-dependent inter-band hopping, is parametrized by the out-of-plane and in-plane coupling constants $\alpha_1$ and $\alpha_2$, respectively. 

The parameters that characterize the tight-binding Hamiltonian in Eq. (\ref{HTIlatt}) can be expressed in terms of the parameters of the continuum theory given by Eq. (\ref{HTI}) by imposing the condition that the two theories coincide in the limit ${\bm k}\rightarrow 0$. Explicitly, we have
\begin{eqnarray}
\epsilon_{\pm} &=& C_0\pm M_0 + \frac{12}{c^2}(C_1\pm M_1) +\frac{4}{a^2}(C_2 \pm M_2), \nonumber \\
t_{1\pm} &=& -\frac{3}{c^2}(C_1\pm M_1), \nonumber \\
t_{2\pm} &=& \frac{1}{c^2}(C_1\pm M_1) - \frac{2}{3a^2}(C_2\pm M_2), \label{tij} \\
\alpha_1 &=& \frac{3}{2} \frac{a A_1}{c^2},  ~~~~~~~~~~\alpha_2=\frac{A_2}{3a}-\frac{aA_1}{2c^2}. \nonumber
\end{eqnarray}
We emphasize that calculating the parameters of the lattice model using  Eq. (\ref{tij}) does not guarantee the correctness of the theory away from the $\Gamma$ point. In particular, one has to ensure that no spurious low-energy states (e.g., gapless states)  are present anywhere inside the Brillouin zone. In addition, the continuum theory itself becomes highly inaccurate away from ${\bm k}=0$. This may not be a problem if we study the low-energy physics of a 3D bulk system, but it becomes critical in the presence of confinement, e.g., in low-dimensional structures. For example, if we consider a slab geometry (i.e. a TI system with a finite number of quintuple layers parallel to the $x-y$ plane), the spacing between the confinement-induced sub-bands is determined by the dispersion along the $k_z$ direction (i.e. the $\Gamma \rightarrow Z$ direction  in the Brillouin zone). Band structure calculations\cite{Zhang2009,Zhang2010a,Liu2010} show that the corresponding bandwidths for the valence and conduction bands are of the order $0.2$eV and $0.7$eV, respectively. The parameters of the lattice model have to be optimized to capture this property. This is illustrated in Fig. \ref{Fig1}, which shows a comparison between the spectrum of the 3D lattice model and that of the continuum model with parameters for Bi$_2$Se$_3$ taken from Ref. \onlinecite{Liu2010}. 

To model a TI film, we confine the system in the $z$ direction by considering a finite number $N_z$ of staked triangular sublattices (i.e., $N_z$ quintuple layers).  A typical low-energy spectrum for a system described by the Hamiltonian in Eq. (\ref{HTIlatt}) in this slab geometry is shown in Fig. \ref{Fig2} (top panel). Notice the characteristic Dirac cone associated with the topologically-protected gapless surface states. 
Finite-size effects due to the hybridization of the surface states corresponding to the opposite surfaces of the slab become significant in very thin films consisting of fewer than about six quintuple layers.\cite{Linder2009,Liu2010a,Lu2010}
A nanoribbon model is obtained by further confining the system in the $x-y$ plane. In this work, we focus on  infinitely long nanoribbons obtained by cutting a TI slab along the $y$ direction, i.e. we consider systems with a finite number $N_x$ of layers oriented perpendicular to the $x$ direction (see the inset of Fig. \ref{Fig1}).  
Finally, we note that, in general, the Hamiltonian (\ref{HTIlatt}) has to be supplemented by a term $H_{\rm V}$ that accounts for local contributions due to disorder, applied gate potentials, or proximity-induced bias potentials, e.g., substrate-induced potentials.\cite{Zhang2010} For convenience, we also include in $H_{\rm V}$ the contribution due to the TI chemical potential $\mu_{TI}$.  The term that accounts for these local contributions has the form
\begin{equation}
H_{\rm V}= \sum_{i, \lambda} [V(i)-\mu_{TI}] c_{i\lambda}^{\dagger}c_{i\lambda}, \label{Hv}
\end{equation} 
where $V(i)$ is a function of position that captures the effect of disorder and the effects generated by gate-  and proximity-induced bias potentials.

\vspace{-3mm}

\subsection{Applied magnetic field} \label{SecIIB}

The applied magnetic field represents one of the key elements of both the semiconductor (SM) wire\cite{Lutchyn2010,Oreg2010} and the TI wire\cite{Cook2011,Cook2012} proposals for realizing Majorana fermions. The role of the magnetic field is to ensure that the quasi-1D system has only one pair of Fermi points (or, more generally, an odd number of pairs) -- the required condition for realizing a topological SC phase -- by opening a gap in the spectrum near $k=0$. We note that, in general, the confinement-induced bands are degenerate at $k=0$ and non-degenerate at finite k-vectors. The $k=0$ degeneracy is protected by time-reversal symmetry, which can be destroyed by applying a magnetic field.  Although the basic role of the applied magnetic field is the same in both types of Majorana structures, there are certain differences stemming from the fact that in SM wires the opening of the gap is mainly due to the Zeeman splitting of the spin sub-bands, while in TI wires the orbital effect is dominant. Ultimately, these differences have quantitative implications, most notably concerning the accessibility of the topological SC phase without requiring fine tunning of the chemical potential or the use of strong magnetic fields.\cite{Stanescu2013,Cook2011}

\begin{figure}[tbp]
\begin{center}
\includegraphics[width=0.48\textwidth]{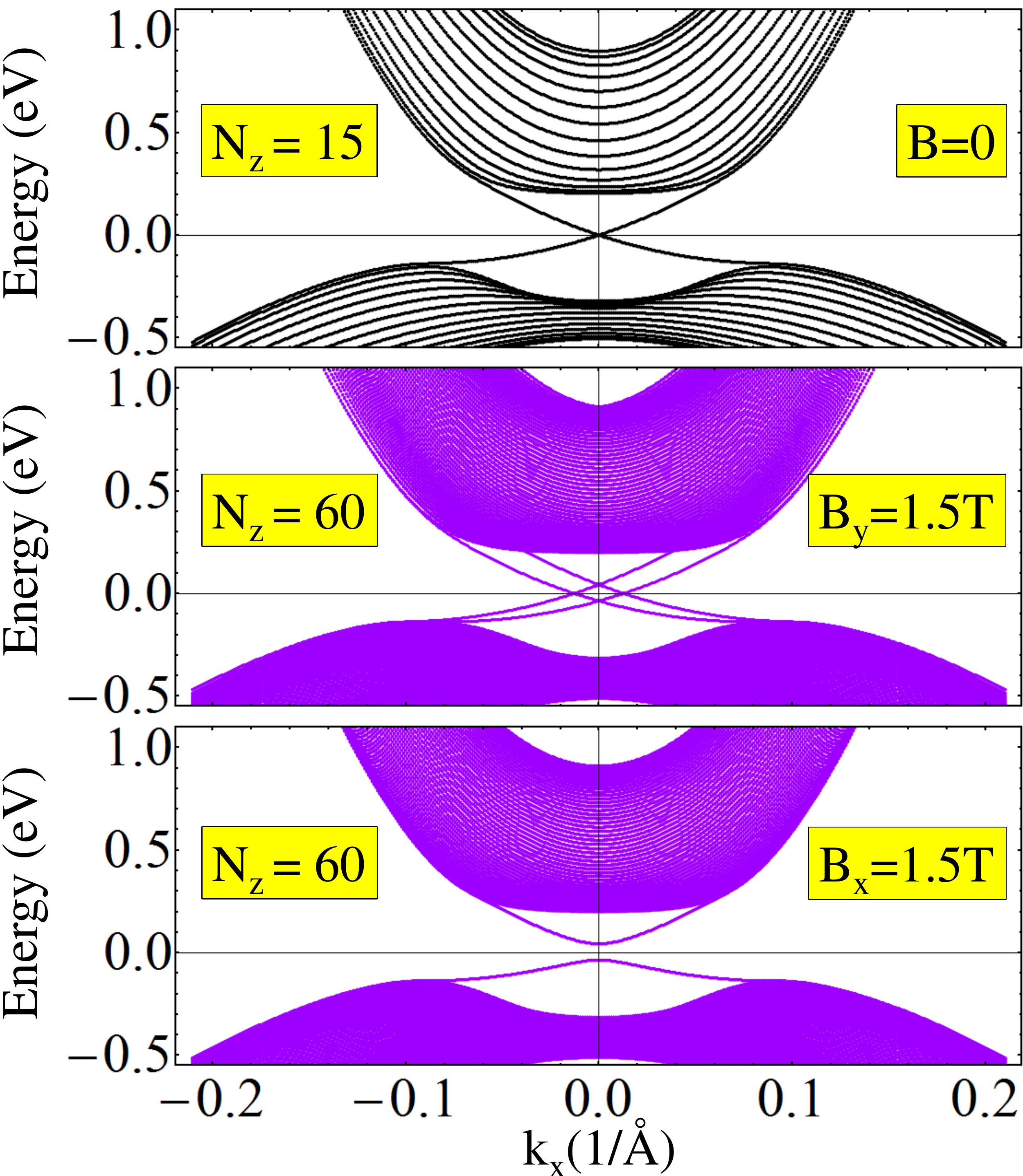}
\vspace{-5mm}
\end{center}
\caption{(Color online) Spectrum of a TI slab with $N_z$ quintuple layers staked along the $z$ direction. The Dirac-cone-like feature (top panel) corresponds to the topologically-protected gapless surface states.  Applying a magnetic field parallel to the slab (middle and bottom panels) removes the double degeneracy of the surface states and shifts the corresponding Dirac cones.  The magnitude of this change depends on the thickness of the slab (i.e. on $N_z$), as it is dominated by the orbital effect (see main text). The effective g factors corresponding to the Zeeman in Eq. (\ref{Hz}) are\cite{Liu2010} $(g_{+z}, g_{-z}, g_{+\parallel}, g_{-\parallel})= (-25.4, -4.12, 4.1, 4.8)$.}
\vspace{-5mm}
\label{Fig2}
\end{figure}

The applied magnetic field modifies the TI spectrum through orbital and Zeeman effects. The orbital effects are incorporated through the Peierls substitution $\gamma_{ij}\rightarrow\gamma_{ij}\exp\left[-\frac{2\pi i}{\Phi_0} \int_{{\bm r}_i}^{{\bm r}_j} {\bm A}\cdot d{\bm l}\right]$, where $\Phi_0=h/2e$ is the magnetic flux quantum and  $\gamma_{ij}$ represent the matrix elements $t_{ij}^{(\lambda)}$ or $\alpha_{ij}$ from Eq. (\ref{HTIlatt}). The Zeeman contribution is captured by the Hamiltonian
\begin{eqnarray}
H_Z = \sum_{i}\sum_{\nu, \nu^\prime}c_{i\nu}^\dagger[{\cal H}_Z]_{\nu,  \nu^\prime} c_{i \nu^\prime},  \label{Hz} 
\end{eqnarray}
with 
\begin{eqnarray}
{\cal H}_Z = \frac{\mu_B}{2}\left(   
\begin{array}{cccc}
g_{+z}B_z & 0 & g_{+\parallel}B_- & 0 \\
0 & g_{-z}B_z & 0 & g_{-\parallel}B_-  \\
g_{+\parallel}B_+ & 0 & -g_{+z}B_z & 0 \\
0 & g_{-\parallel}B_+ & 0 & -g_{-z}B_z                  
\end{array}\right),       
\end{eqnarray}
where $\nu=(\lambda, \sigma)$ and $\nu^\prime=(\lambda^\prime, \sigma^\prime)$, $\mu_B$ is the Bohr magneton, $B_{\pm}=B_x\pm B_y$ and $B_z$ are the components of the magnetic field and $g_{\pm z}$, $g_{\pm \parallel}$ are effective g factors derived from ${\bm k}\cdot{\bm p}$ theory\cite{Liu2010}. To illustrate the effect of the magnetic field, we calculate the spectrum of a TI slab in the presence of a magnetic field applied along a direction parallel to the slab. The results are shown in Fig. \ref{Fig2} (middle and bottom panels). We note that, for this orientation of the magnetic field, the surface states remain gapless, but the corresponding Dirac cones are shifted along the in-plane direction perpendicular to the field. Another important observation is that the Zeeman splitting is negligible compared with the orbital effect. For the 60 quintuple layer slab shown in Fig. \ref{Fig2}, the relative contribution of the Zeeman effect is about $0.5\%$ of the change in energy for the surface states. This is also the case in TI nanoribbons\cite{Cook2011,Cook2012} and represents one of the differences between the TI-based Majorana structures and their semiconductor-based counterparts,\cite{Lutchyn2010,Oreg2010,Mourik2012} where Zeeman splitting is the dominant effect.   

\vspace{-3mm}

\subsection{Superconducting proximity effect} \label{SecIIC}

The pairing correlations required to engineer the topological superconducting state are obtained by coupling the TI nanoribon to an ordinary s-wave superconductor. The SC component of the hybrid system is modeled at the mean-field level using a simple tight-binding Hamiltonian characterized by a local pairing amplitude $\Delta_0$. Explicitly, we have
\begin{equation}
H_{\rm SC} = \sum_{i, j, \sigma}(t_{ij}^{sc} - \mu_{sc}\delta_{ij})a_{i\sigma}^\dagger a_{j \sigma} + \Delta_0\sum_{i}(a_{i\uparrow}^\dagger a_{i\downarrow}^\dagger + a_{i\downarrow}a_{i\uparrow}), 
\end{equation}
where $a_{i\sigma}^\dagger$ is the creation operator for states with spin $\sigma$ localized near site $i$,  and $\mu_{sc}$ is the chemical potential of the SC. For concreteness, we assume nearest-neighbor hopping, i.e. $t_{ij}^{sc}=-t^{sc}$ if $i$ and $j$ are nearest-neighbors and zero otherwise. In the numerical calculations we use $\Delta_0=1.5$meV.
The coupling between the TI described by the lattice Hamiltonian (\ref{HTIlatt}) and the SC is captured by the coupling term
\begin{equation}
H_{\rm TI-SC} =  \sum_{i_0, j_0}\sum_{\lambda\sigma}(\tilde{t}_{\lambda\sigma}c_{i_0\lambda\sigma}^\dagger a_{j_0\sigma} + h.c.),
\end{equation}
where $i_0$ and $j_0$ are sites at the TI-SC interface inside the TI and SC regions, respectively, and $\tilde{t}_{\lambda\sigma}$ are spin- and orbital-dependent coupling constants. For simplicity we assume lattice matching across the interface. To understand the structure of the coupling matrix, we note that the TI molecular orbitals given by Eq. (\ref{states}), which are superpositions of p-type orbitals, contain $p_z$ components that couple strongly with SC states across the interface and $p_\pm$ components, i.e. p-orbitals oriented parallel to the interface, that are weakly coupled to the SC because of their orientation and because of the sign change between the lobes. In the following, we assume that only the states $|\lambda, p_z, \sigma\rangle$ have non-vanishing matrix elements with the SC states. In addition, we assume that hopping across the interface preserves the spin and is independent of the spin orientation. Consequently, the coupling matrix element between a state $|\lambda, \sigma\rangle$ given by equation (\ref{states}) and a SC state with spin $\sigma $ is
\begin{equation}
\tilde{t}_{\lambda\sigma}=\delta_{\sigma\sigma^\prime}\left(
\begin{array}{cc}
\tilde{t}_+ & 0 \\
\tilde{t}_- & 0 \\
0 & \tilde{t}_+ \\
0 &\tilde{t}_- 
\end{array}\right)_{\lambda\sigma, \sigma^\prime}.
\end{equation}
The relative strength of $\tilde{t}_+$ and  $\tilde{t}_-$  is determined by the coefficients $u_\pm$ from Eq. (\ref{states}). However, to better understand the possible effect of varying these mixing coefficients (e.g., by changing the relative strength of the spin-orbit coupling),  we treat $\xi=\tilde{t}_-/\tilde{t}_+$ as an independent model parameter. 

The effective low-energy theory for the TI subsystem can be obtained using the Green's function formalism by integrating out the SC degrees of freedom.\cite{Stanescu2010,Stanescu2013} After integration, the effect of the proximity-coupled SC is captured by a surface self-energy that supplements the bare TI Green's function. Explicitly, we have
\begin{equation}
\Sigma_{\lambda \sigma, \lambda^\prime \sigma^\prime}(\omega; i_0, i_0^\prime) =\tilde{t}_{\lambda\sigma}G_{\sigma, \sigma^\prime}^{SC} (\omega; j_0, j_0^\prime)\tilde{t}_{\lambda^\prime\sigma^\prime},                        \label{sigma}
\end{equation}
where $(i_0, j_0)$ and $(i_0^\prime, j_0^\prime)$ are pairs of nearest neighbor sites on the two sides of the TI-SC interface, and $G_{\sigma, \sigma^\prime}^{SC}$ is the superconductor Green's function. We note that $G^{SC}(\omega; j_0, j_0^\prime)$ is a short-range function of $|{\bm r}_{j_0}-{\bm r}_{j_0^\prime}|$, i.e. it decreases rapidly with the distance between the two points on the SC boundary. Since we are interested in the low-energy long-wavelength physics of the TI subsystem, we can approximate the surface SC Green's function by the local contribution  $G^{SC}(\omega; j_0)= \delta_{ j_0 j_0^\prime} G^{SC}(\omega; j_0, j_0^\prime)$. For a planar surface, this local contribution is independent of position, and we have 
\begin{equation}
G^{SC}(\omega) = -\nu_F\left[\frac{\omega +\Delta_0\sigma_y\tau_y}{\sqrt{\Delta_0^2-\omega^2}}+\zeta \tau_z\right], \label{Gsc}
\end{equation}
with $\sigma_\alpha$ and $\tau_\alpha$ being Pauli matrices in the spin and particle-hole spaces, respectively, and $\nu_F=\frac{1}{|t^{sc}|}\sqrt{1-\left(1-\frac{\mu_{sc}}{2t^{sc}}\right)^2}$ is the (surface) density of states at the Fermi energy for the bulk SC in the normal phase. The last term in Eq. (\ref{Gsc}) represents a proximity-induced interface bias potential and will be incorporated into the local potential term $H_V$ given by Eq. (\ref{Hv}).  We note that Eq. (\ref{Gsc}) holds for $|\omega| <\Delta_0$. When the frequency is comparable to the bulk SC pairing potential $\Delta_0$, dynamical effects become important and cannot be neglected. In this work, however, we focus on the low-energy regime characterized by $|\omega|\ll \Delta_0$ and treat the superconducting proximity effect within the static approximation  $\sqrt{\Delta_0^2-\omega^2}\approx \Delta_0$. We note that this approximation, rather than being too restrictive, provides accurate results for the low-energy spectrum for frequencies up to $0.4\Delta_0$.\cite{Stanescu2011} In general, the low-energy spectrum of the proximity-coupled TI subsystem can be obtained by solving the Bogoliubov-de Gennes (BdG) equation ${\rm det}\left[G_{\rm TI}^{-1}(\omega)\right] = 0$, where the TI Green's function is 
\begin{equation}
G_{\rm TI}(\omega) = \left[\omega - \xbar{H}_{\rm TI} -\xbar{H}_{\rm V}-\xbar{H}_{\rm Z}-\Sigma(\omega)\right]^{-1},  \label{GTI1}
\end{equation}
with
\begin{equation}
\xbar{H}_{\rm X} = \left(
\begin{array}{cc}
H_{\rm X} & 0 \\
0 & -H_{\rm X}^{T}
\end{array}\right),
\end{equation}
 where X $=$ TI, V, Z, with the corresponding Hamiltonians given by equations  (\ref{HTIlatt}), (\ref{Hv}), and  (\ref{Hz}), respectively, and the self-energy $\Sigma(\omega)$ is given by Eq. (\ref{sigma}). Using Eq. (\ref{Gsc}) within the static approximation, one can explicitly write the proximity-induced interface self-energy as 
\begin{equation}
\Sigma(\omega) = - \frac{\gamma}{\Delta_0}\left[\omega M_0(\xi) + i\tau_y \Delta_0 M_1(\xi)\right] \hat{K}, \label{sigma1}
\end{equation}
where $\gamma = (\tilde{t}_+^2+\tilde{t}_-^2)\nu_F$ is the effective coupling strength at the TI-SC interface and $\hat{K}$ is a matrix  that has non-vanishing local contributions only at the interface, i.e. $\hat{K}_{ij}=\delta_{ij}$ if $i=i_0$ is an interface site and $\hat{K}_{ij}=0$ otherwise. The matrices $M_0$ and $M_1$ represent proximity-induced normal and anomalous self-energy corrections, respectively. We note that, to simplify notation,  here and throughout the paper we do not write explicitly the unit matrices that multiply certain terms -- e.g.,  the unit matrix in the particle-hole space multiplying the $M_0$ term in Eq. (\ref{sigma1}) -- and we omit direct product symbols, e.g., $\tau_y \otimes M_1 \otimes \hat{K} \rightarrow  \tau_y M_1 \hat{K}$.
The matrices $M_\alpha(\xi)$, which depend on the relative strength of the coupling matrix elements, $\xi=\tilde{t}_-/\tilde{t}_+$, capture the orbital dependence of the TI-SC coupling and have the following expressions:
\begin{eqnarray}
M_0(\xi)&=&\frac{1}{1+\xi^2}\left(
\begin{array}{cccc}
1 & \xi & 0 & 0 \\
\xi & \xi^2 & 0 & 0 \\
0 & 0 & 1 & \xi \\
0 & 0 & \xi & \xi^2
\end{array}\right), \nonumber \\
M_1(\xi)&=&\frac{1}{1+\xi^2}\left(
\begin{array}{cccc}
0 & 0 & -1 & -\xi  \\
0 & 0 & -\xi & -\xi^2 \\
1 & \xi & 0 & 0 \\
\xi & \xi^2 & 0 & 0 
\end{array}\right).
\end{eqnarray}
Next, we notice that the frequency-dependent term in Eq. (\ref{GTI1}) has the form $\omega Q$, with $Q=1+\frac{\gamma}{\Delta_0} M_0(\xi) \hat{K}$ being a positive definite matrix that can be mapped into the unit matrix using the transformation
\begin{equation}
\widetilde{Z}\left[1+\frac{\gamma}{\Delta_0} M_0(\xi) \hat{K}\right]\widetilde{Z}^T=1,
\end{equation}
where the renormalization matrix has the form
\begin{eqnarray}
\widetilde{Z} &=& 1-\hat{K}  \\
&+& \frac{1}{\sqrt{1+\xi^2}}\left(
\begin{array}{cccc}
\beta &\beta \xi & 0 &0 \\
-\xi & 1 & 0 & 0 \\
0 & 0 & \beta &\beta \xi  \\
0 & 0 & -\xi & 1
\end{array}\right)\hat{K}, \nonumber
\end{eqnarray}
where $\beta=1/\sqrt{1+\frac{\gamma}{\Delta_0}}$. 
Consequently, we can rewrite the BdG equation as  ${\rm det}\left[\omega -H_{\rm eff}\right]=0$, which reduces the BdG problem to finding the eigenvalues of an effective low-energy Hamiltonian. Putting together all these elements, we conclude that the low-energy physics of a low-dimensional TI system (e.g., TI film or nanoribbon) proximity-coupled to an ordinary s-wave superconductor is described by the effective BdG Hamiltonian
\begin{equation}
H_{\rm eff} = \widetilde{Z}\left[\xbar{H}_{\rm TI} +\xbar{H}_{\rm V}+\xbar{H}_{\rm Z}\right]\widetilde{Z}^T + i\tau_y\left(
\begin{array}{cc}
\Delta_{\rm ind} & 0 \\
0 & 0
\end{array}\right) \hat{K}, \label{Heff}
\end{equation}  
where the proximity-induced pairing potential is $\Delta_{\rm ind} = \gamma\Delta_0/(\gamma + \Delta_0)$. We note that, in addition to the pairing term in Eq. (\ref{Heff}), the superconducting proximity effect results in a renormalization of the low-energy spectrum through the renormalization matrix $\widetilde{Z}$. We emphasize that both of these contributions are nonzero only on lattice sites located at the TI-SC interface. In the remainder of this work, we study the low-energy physics of TI-SC hybrid structures by numerically solving the eigenvalue problem for the effective Hamiltonian given by Eq. (\ref{Heff}).   

\begin{figure}[tbp]
\begin{center}
\includegraphics[width=0.48\textwidth]{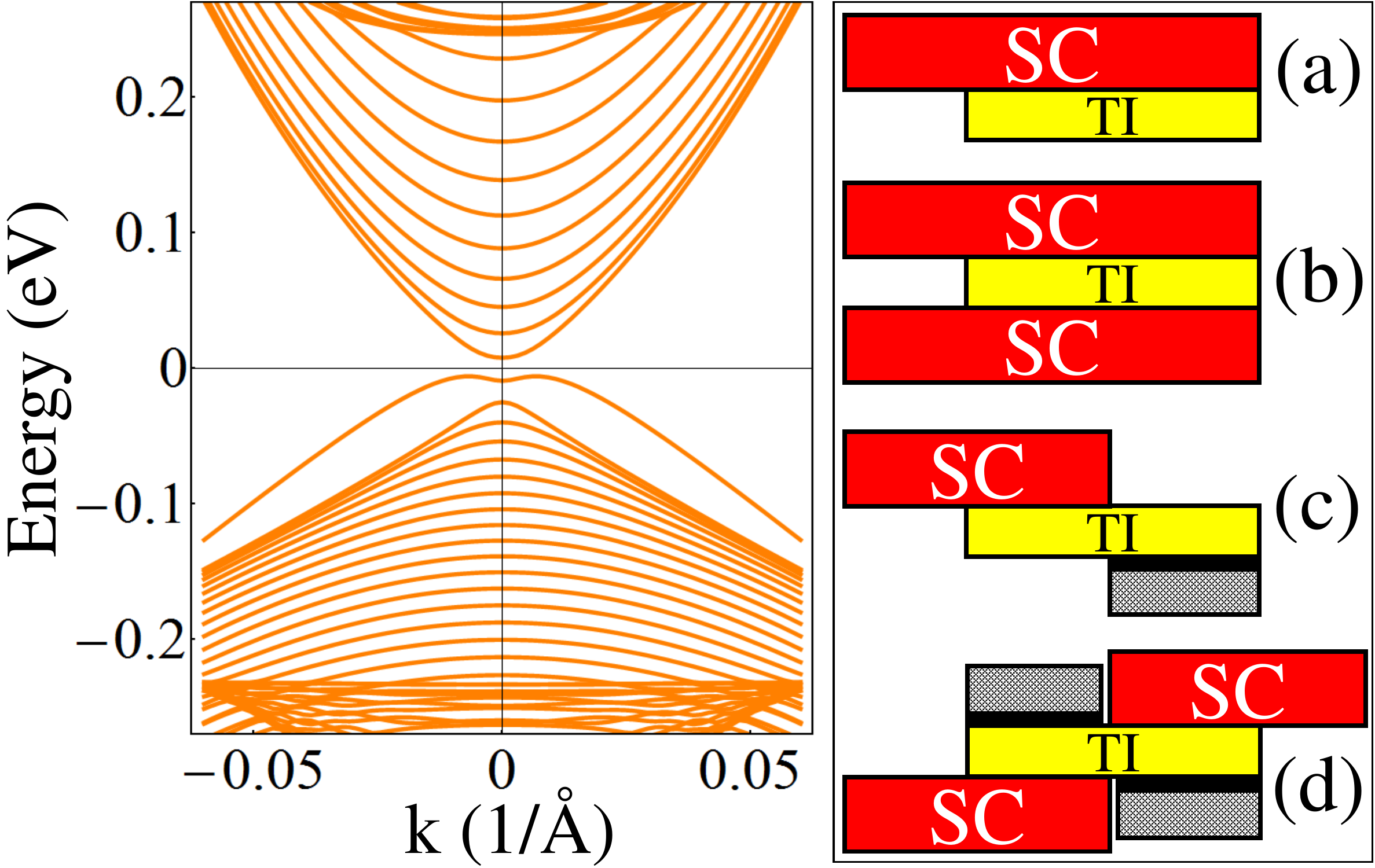}
\vspace{-5mm}
\end{center}
\caption{(Color online) Low-energy spectrum of a TI nanoribbon with rectangular cross section $L_x\times L_z=103\times 9$~nm (left) and schematic representation of typical proximity-coupled nanostructures studied in this work (right). All the bands shown in the left panel are double degenerate. The bands with energies within the bulk gap result from confining the TI surface states (see Fig. \ref{Fig2}) along the $x$ direction. Notice the small gap near zero energy. The cross sectional views in the right panel show the main components of a TI-based hybrid structure, including back/top gates (structures c and d) for controlling the chemical potential. Note that interface-induced bias potentials generated by the proximity-coupled bulk SC or  the substrate\cite{Zhang2010} (not shown) are always present in non-symmetric structures (e.g., structure a), even in the absence of applied gate potentials.}
\vspace{-3mm}
\label{Fig3}
\end{figure}

\section{Numerical study of superconducting TI nanoribbons} \label{SecIII}

In this section we present the results of numerical calculations based on the model described above. We focus on three aspects of the low-energy physics in proximity-coupled TI nanoribbons: A) Understanding the basic structure of the energy spectrum and of its dependence on control parameters, such as applied magnetic fields and bias potentials. We also focus on determining the real-space structure of low-energy states under various conditions, as the amplitudes of these states near the interface control the strength of the TI-SC proximity effect. B) Mapping out the phase diagram and determining the dependence of the phase boundaries on relevant model and control parameters. C) Calculating the magnitude of the proximity-induced quasiparticle gap under experimentally-relevant conditions and identifying possible directions for maximizing this gap in the topological SC phase. Since the stability of the topological SC phase and, ultimately, the robustness of the zero-energy Majorana bound states hosted by TI nanoribbons that support topological superconductivity depend critically on the magnitude of the induced gap, this analysis  has direct practical implications.

\begin{figure}[tbp]
\begin{center}
\includegraphics[width=0.49\textwidth]{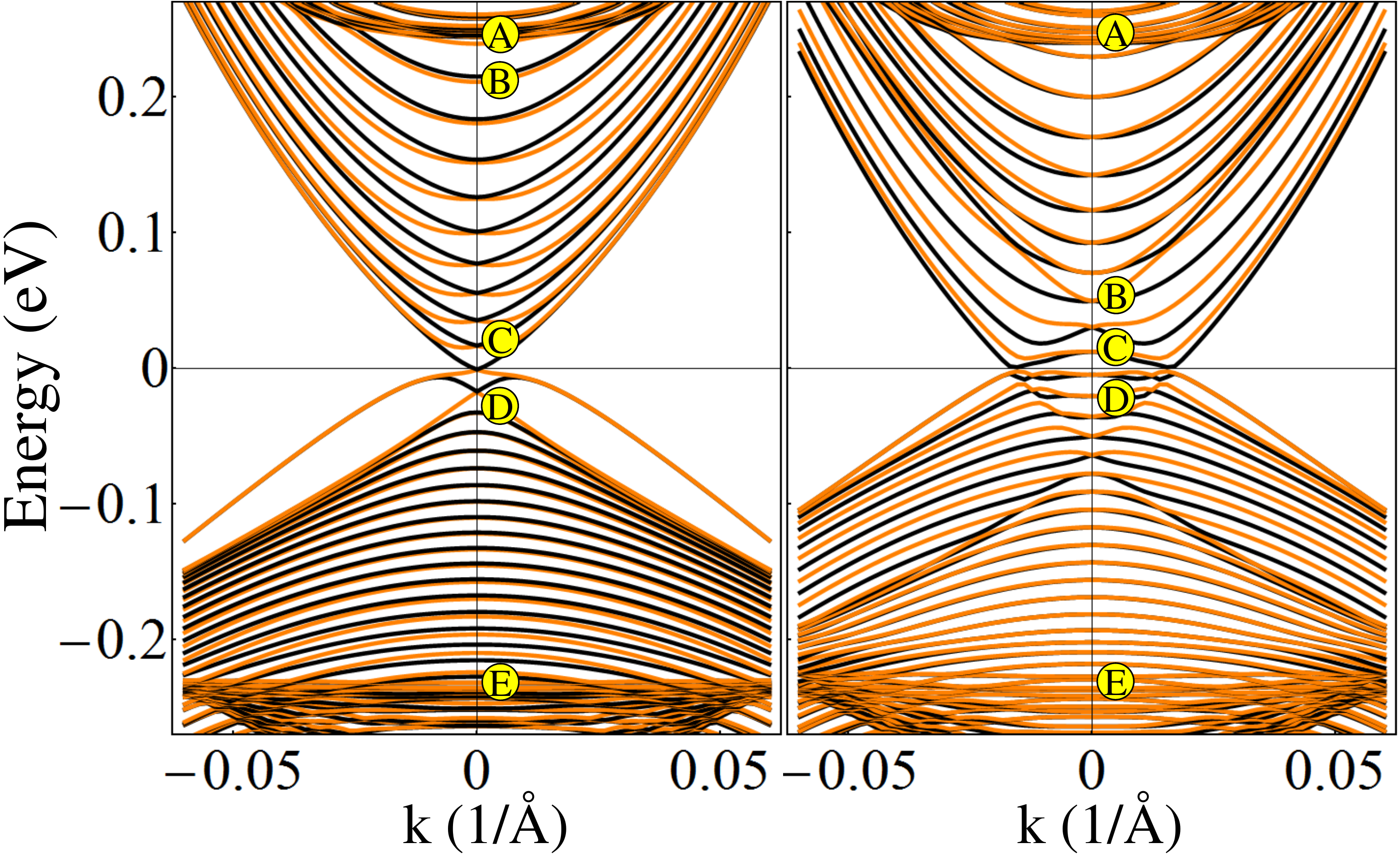}
\vspace{-5mm}
\end{center}
\caption{(Color online) {\em Left}: Low-energy nanoribbon spectrum in the presence of an external magnetic field parallel to the wire. The total flux through the ribbon is $\Phi=0.583\Phi_0$, which realizes the $k=0$ degeneracy condition for the lowest energy bands (see main text). Note that for this value of $\Phi$ the higher energy bands are not exactly degenerate at $k=0$. {\em Right}: Low-energy spectrum in the presence of a bias potential with $V(i)$ given by Eq. (\ref{Vi}) and $V_{\rm max}=0.05$~eV. The spatial profiles of the states marked by letters (A--E) are shown in Fig. \ref{Fig5} (left panel) and Fig. \ref{Fig6} (right).}  
\vspace{-3mm}
\label{Fig4}
\end{figure}

\vspace{-2mm}

\subsection{Spectrum and low-energy states} \label{SecIIIA}

\subsubsection{Normal state nanoribbons} \label{SecIIIA1}

We start our analysis by discussing the main characteristics of the low-energy spectrum of a TI nanoribbon in the normal state, i.e., in the absence of the proximity-coupled superconductor, and the qualitative changes induced by external magnetic fields and bias potentials. We consider infinitely long  nanoribbons ($L_y\rightarrow\infty$) obtained by cutting a TI slab of thickness $L_z=N_z c/3$ into stripes of thickness $L_x=N_x a$ (see the inset of Fig. \ref{Fig1}). A typical low-energy spectrum is shown in Fig. \ref{Fig3} (left panel). We emphasize that this spectrum corresponds to a free-standing ribbon and is characterized by doubly degenerate bands. However, TI-based hybrid structures (see Fig. \ref{Fig3}, right panel) contain interfaces  that, typically, lower the symmetry of the system and generate bias potentials\cite{Zhang2010} that remove the band degeneracy. Also, we note that the spectrum is characterized by a small gap near zero energy. This is always the case for TI nanoribbons and nanowires (finite or infinitely long) and reflects the fact that the quasi-1D system is in a topologically {\em trivial} phase. In other words, systems described by class AII (symplectic) Hamiltonians, such as that in Eq. (\ref{HTIlatt}), can support topological phases characterized by a ${\mathbb{Z}}_2$ topological invariant in two and three dimensions, but only topologically trivial phases exist in one dimension.\cite{Schnyder2008}  In this work, the term 'topological insulator nanoribbon' designates a quasi-1D system made of materials that support 3D topological phases. 

\begin{figure}[tbp]
\begin{center}
\includegraphics[width=0.48\textwidth]{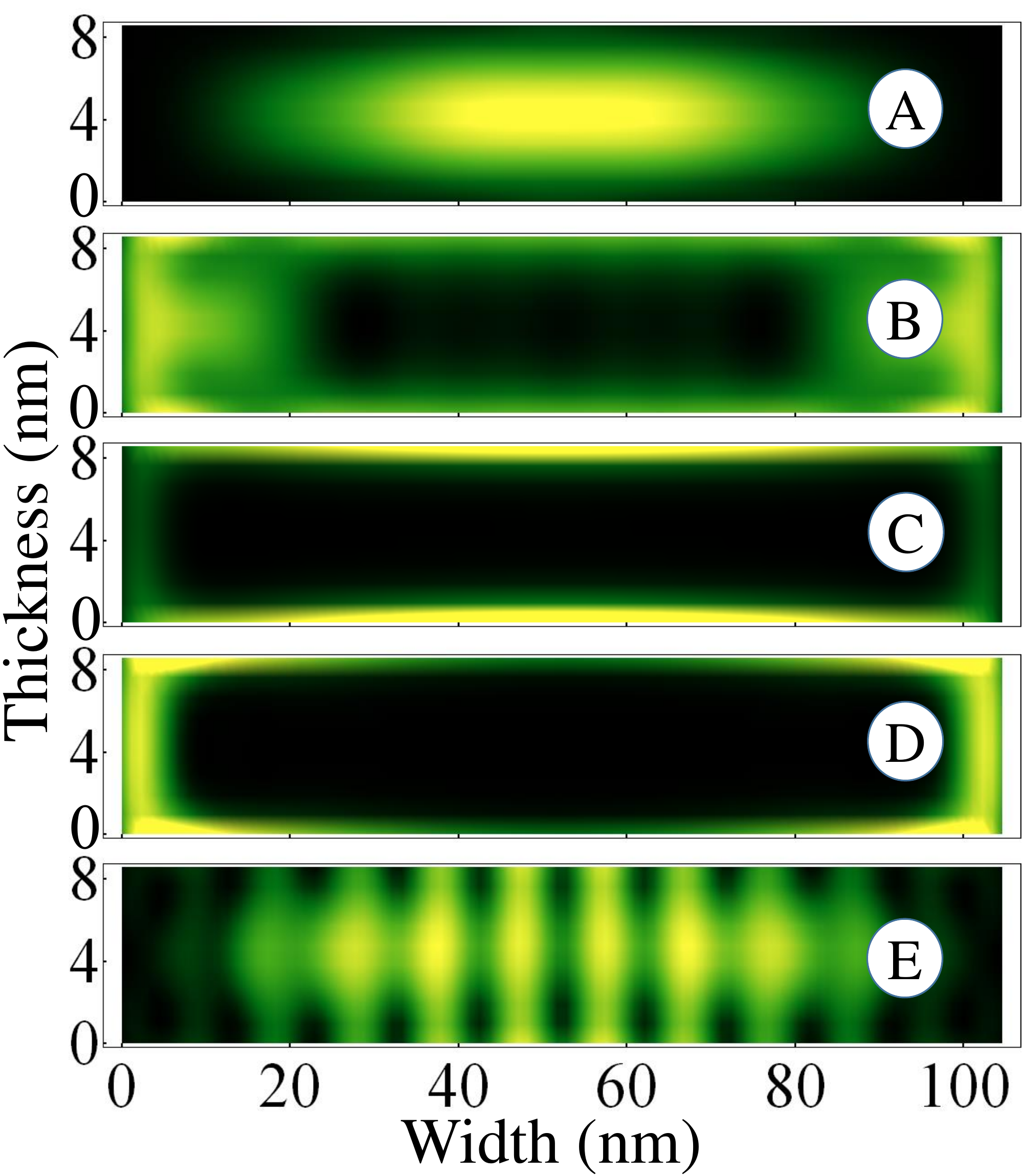}
\vspace{-5mm}
\end{center}
\caption{(Color online) Typical transverse profiles $|\psi_n(x, z)|^2$ for low-energy states in TI nanoribbons. Yellow (light gray) regions correspond to the maxima of the wave functions. Panel A shows a bulk-type state with energy $E_{\rm A}=0.24$~eV, while panel E corresponds to a valence bulk-type band with $E_{\rm E}=-0.22$~eV (see Fig. \ref{Fig4}). The surface-type bands with energies within the bulk gap  (see Fig. \ref{Fig4}) are characterized by wave functions with maxima near the boundaries of the system (panels B, C, and D).}
\vspace{-3mm}
\label{Fig5}
\end{figure}

The proximity-coupled ribbons are driven into a topological SC phase by applying a magnetic field parallel to the wire, i.e. along the $y$ direction. The magnetic field removes the band degeneracy, as already suggested by the behavior of the surface-state Dirac cone in the slab geometry calculation shown in Fig. \ref{Fig2}. Quantitatively, most of the change in the spectrum is due to the orbital effect, as explained above (see Sec. \ref{SecIIB}).  When the magnetic flux through the ribbon, $\Phi= B L_x L_z$,  is approximately equal to half (or a half-integer multiple) of $\Phi_0=h/2e$, the bands again become degenerate at $k=0$. However, as shown in Fig. \ref{Fig4} (left panel), the corresponding spectrum has the remarkable property that the number of pairs of Fermi points $\{-k_F^{(n)}, k_F^{(n)}\}$ associated with an arbitrary value of the chemical potential within the bulk TI gap is odd. Since this is precisely the necessary condition for realizing the topological SC phase that supports Majorana zero-energy modes, the TI-based hybrid structures open the possibility of being able to realize Majorana quasiparticles without fine tunning the chemical potential.\cite{Cook2011,Cook2012}  
We note a finite-size effect,  clearly visible in thin wires, that results in the $k=0$ degeneracy condition being realized for different, band-dependent values of $\Phi$ slightly larger than  $0.5\Phi_0$ (see Fig. \ref{Fig4}).  

\begin{figure}[tbp]
\begin{center}
\includegraphics[width=0.48\textwidth]{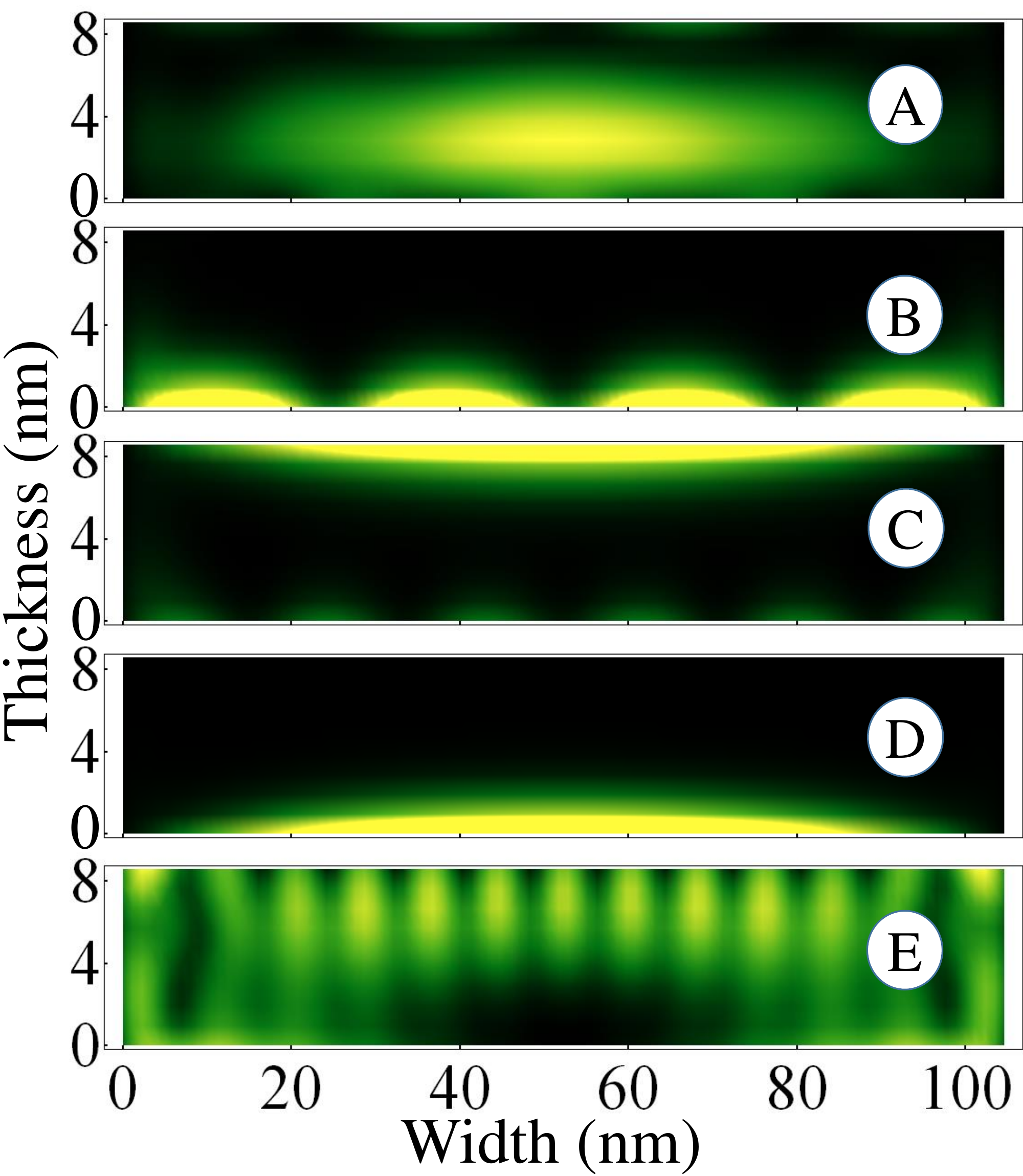}
\vspace{-5mm}
\end{center}
\caption{(Color online) The effect of a bias potential on the spatial distribution of low-energy wave functions. While the bulk-type states (A and E) are only slightly distorted by the applied potential, the surface-type states become practically localized near the top (panel B) or the bottom (panels C and D) surfaces  (see Fig. \ref{Fig4}). Proximity-coupling the TI ribbon to a bulk SC placed on the top (see Fig. \ref{Fig3}, structure a) will result in a large proximity-induced gap in band B, but vanishingly small gaps in bands C and D. }
\vspace{-3mm}
\label{Fig6}
\end{figure}

To determine the effect of a bias potential on the low-energy spectrum, we consider  in Eq. (\ref{Hv}) a linear position-dependent function of the form
\begin{equation}
V(i) = \frac{V_{\rm max}}{2}\left(i_z - \frac{N_z+1}{2}\right), \label{Vi}
\end{equation}   
where $i=(i_x, i_y, i_z)$ gives the position of a lattice site and $N_z$ denotes the number of quintuple layers.  We note that the results are basically determined by the potential difference $V_{\rm max}$ between the top and bottom surfaces and depend weakly on the details of $V(i)$. This is due to the fact that the low-energy wave functions have most of their spectral weight in the vicinity of the boundaries, as we will show below.  Consequently, the bias potential (\ref{Vi}) will have an effect similar to that of an interface-induced potential in a system such as structure (a) (see Fig. \ref{Fig3}).  As shown in Fig. \ref{Fig4} (right panel), the bias potential removes the double degeneracy of the bands at finite k-vectors. The degeneracy at $k=0$ is preserved by time-reversal symmetry and, as a consequence, the system is characterized by an even number of pairs of Fermi points for any value of the chemical potential. To realize the topological condition (odd number of Fermi pairs), one has to apply a magnetic field.  Note that for half-integer values of $\Phi/\Phi_0$, the special property discussed above holds in the presence of a bias potential.   

Before discussing the spectrum of proximity-coupled nanoribbons, it is useful to understand the spatial structure of the low-energy states. More specifically, we are interested in the amplitude $|\psi_n(i_0)|^2=\sum_{\lambda,\sigma}|\psi_{n\lambda\sigma}(i_0)|^2$ of the low-energy wave functions near the boundaries of the nanoribbon, particularly near the surfaces that will be interfaced with superconductors. Practically, these interface amplitudes determine the strength of the superconducting proximity effect\cite{Stanescu2011,Stanescu2013} by controlling the coupling to the proximity-induced surface terms in the effective Hamiltonian (\ref{Heff}). Typical transverse profiles $|\psi_n(i_x, i_z)|^2$ corresponding to several low-energy bands are shown in Fig. \ref{Fig5} and Fig. \ref{Fig6}. We note that the wave functions  are normalized, $\sum_{i_x,i_z}|\psi_n(i_x, i_z)|^2=1$, and that the profiles depend very weakly on k for non-degenerate bands.  As shown in Fig. \ref{Fig5}, there are two types of states: bulk-type states (A and E) with energies corresponding to the bottom (top) of the conduction (valence) band in Fig. \ref{Fig1} and Fig. \ref{Fig2} and surface-type states (B, C, and D) with energies within the bulk TI gap. Our first important result, shown in Fig. \ref{Fig6}, demonstrates that in the presence of a bias potential, including interface- and substrate-induced potentials in asymmetric structures, the surface-type states become localized near the top or bottom surfaces. Consequently, if the TI-SC interface corresponds to a TI surface with vanishingly small wave function amplitude, the proximity-induced gap of the corresponding band will be extremely small. This feature can potentially negate one of the main advantages of the TI wire-SC hybrid structures, namely the absence of a strong constraint on the value of the chemical potential, as weakly coupled bands have to be avoided.  A more detailed analysis of this problem is presented in Sec. \ref{SecIIIC}.

\begin{figure}[tbp]
\begin{center}
\includegraphics[width=0.48\textwidth]{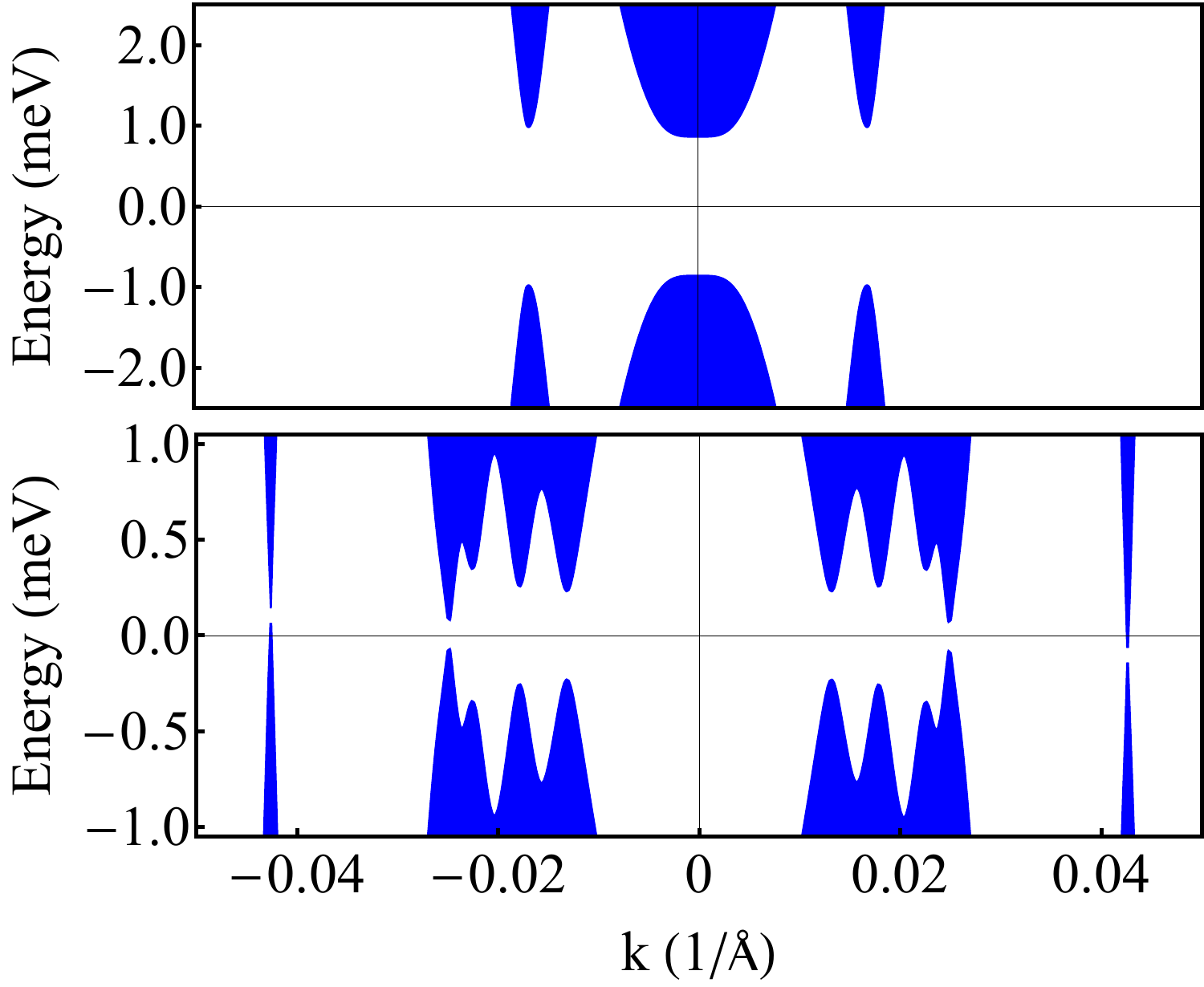}
\vspace{-8mm}
\end{center}
\caption{(Color online) Proximity-induced quasiparticle gap characterizing the low-energy spectrum of the TI nanoribbon - superconductor hybrid  structure (b) from Fig. \ref{Fig3}.  The bulk pair potential is $\Delta_0=1.5$ meV, $\phi_{SC}=0$, the TI-SC coupling strength is $\gamma=4\Delta_0$. and $\xi=0.5$. {\em Top}: Induced gap for $B=0$ and $\mu_{TI}=0.046$ eV. {\em Bottom}: Non-zero magnetic field corresponding to $\Phi=0.8\Phi_0$ and chemical potential $\mu_{TI}=-0.086$ eV. Note that,  in the absence of time-reversal symmetry ($B\neq 0$) the spectrum has particle-hole symmetry, $E(-k)=-E(k)$, but $E(-k)\neq E(k)$. As a result, the SC state corresponding to the lower panel is gapless.}
\vspace{-5mm}
\label{Fig7}
\end{figure}

\vspace{-3mm}

\subsubsection{Superconducting nanoribbons} \label{SecIIIA2}

Let us consider now a TI nanoribbon - superconductor hybrid system, i.e. turn on the coupling across the TI-SC interface, and calculate the low-energy spectrum by solving numerically the eigenvalue problem for the  Bogoliubov de Gennes (BdG) effective Hamiltonian (\ref{Heff}). Here we focus on the main aspect of the superconducting proximity effect, namely the emergence of a proximity-induced quasiparticle gap, while the  dependence of this gap on various relevant parameters (e.g., applied magnetic fields and bias potentials) will be discussed in detail in Sec. \ref{SecIIIC}.  For concreteness, we consider a system corresponding to structure (b) in Fig. \ref{Fig3}, i.e. a nanoribbon sandwiched between two identical SCs, in which the proximity-induced interface bias potential [see the discussion following Eq. (\ref{Gsc})] vanishes by symmetry. In the numerical calculations we typically consider a TI nanoribbon with $L_z=9.5$nm and $L_x=60$nm, unless explicitly stated otherwise. 
Before discussing the results, we note two technical issues that occur when the TI nanoribbon is coupled to two superconductors. First, it is possible to have a phase difference $\phi_{SC}$ between the superconductors, e.g., the order parameters are $\Delta_0$ for the lower SC and $\Delta_0\, e^{i\phi_{SC}}$ for the top SC. This will lead to different values of the proximity-induced pair potential on the two interfaces, i.e. in Eq. (\ref{Heff}) we will have $\Delta_{\rm ind}$ at the bottom interface and $\Delta_{\rm ind}\, e^{i\phi_{SC}}$ at the top interface. The second aspect is more subtle and concerns the incorporation of the orbital effects of the magnetic field through the Peierls substitution (see Sec. \ref{SecIIB}) and the fact that the magnetic field vanishes inside the SCs. Assume that we choose the vector field ${\bm A} = (B z, 0, 0)$ to describe a magnetic field oriented parallel to the wire (i.e. in the $y$ direction). The vector field has to be constant inside the superconductors, because the magnetic field vanishes, so we have ${\bm A} = 0$ for $z<0$ and ${\bm A} = (B L_z, 0, 0)$ for $z>L_z$. For consistency, the nonzero vector field for $z>L_z$ has to be absorbed as an $x$-dependent phase factor in the tight-binding model for the top SC and will generate an additional phase factor for the proximity-induced pair potential on the top surface. Specifically,  the induced pair potential for $i_z=N_z$ becomes $\Delta_{\rm ind}\, e^{i(\phi_{SC} -2 B (i_x-1)a L_z)}$. 

First,  we address the case corresponding to a zero magnetic field and a chemical potential $\mu_{TI}$ placed at the bottom of the second positive-energy band (see Fig. \ref{Fig3}), which corresponds to $k_F^{(1)}\approx 0.017/a$ for the lowest (positive) band and $k_F^{(2)}=0$ for the second band. In the presence of the proximity effect generated by a bulk SC with $\Delta_0=1.5$meV, a quasiparticle gap $\Delta_{qp}\approx 1$meV opens at the Fermi points, as shown in the upper panel of Fig. \ref{Fig7}. Note that, in the absence of an applied magnetic field, the expected induced gap for a certain band $n$ is 
$\Delta_{qp}^{(n)} \approx \Delta_{\rm ind} \sum_{i_0} |\psi_n(i_0)|^2$, where $i_0$ labels the interface sites. For the parameters used in this calculation, $\Delta_{qp}^{(n)} \approx 0.8\Delta_{\rm ind}$, which means that the two relevant bands are characterized by surface-type states similar to state C in Fig. \ref{Fig5} and having about $80\%$ of their weight on the top and bottom surfaces.

Next, we consider a case characterized by a nonzero magnetic field and a chemical potential $\mu_{TI}$ that crosses multiple bands. The results are shown in the lower panel of Fig. \ref{Fig7}.   The quasiparticle gaps that open near the Fermi points corresponding to different bands can have significantly different values, as a result of those bands having different transverse profiles (see Fig. \ref{Fig5}) and, consequently, different wave-function amplitudes at the TI-SC interface.  Most interestingly, the low-energy BdG spectrum has the property  $E(-k)=-E(k)$, as a result of particle-hole symmetry, but, for a given wave-vector, there is no correspondence between positive and negative eigenvalues (i.e. $E_-(k) \neq -E_+(k)$). Consequently, for certain parameters, the induced superconducting state can become gapless, as illustrated in Fig. \ref{Fig7} (bottom panel) by the vanishing of the gap near the largest Fermi point (which corresponds to the topmost negative-energy band in Fig. \ref{Fig3}). When this type of situation occurs in the topological SC phase, the system may still support zero-energy Majorana bound states, but they will not be robust against any type of disorder.

Before we conclude this section, it is instructive to discuss the source of the asymmetry between positive and negative bands and to understand the difference between our model and the models studied by Cook and Franz,\cite{Cook2011} which are characterized by  $E_-(k) = -E_+(k)$. In the long- wavelength limit, the simple effective Hamiltonian of Ref. \onlinecite{Cook2011} can be written as
\begin{equation}
{\mathcal H}_1(k) = (k\sigma_y -\mu_{TI} +m\sigma_z)\tau_z - \Delta \tau_y\sigma_y, \label{H1}
\end{equation}
where $\tau_\alpha$ and $\sigma_\alpha$ are Pauli matrices in the particle-hole and spin spaces, respectively, $m$ represents the Zeeman splitting, and $\Delta$ is the induced pair potential. As expected, the Hamiltonian (\ref{H1}) has particle-hole symmetry, ${\mathcal H}_1(k) =-\tau_x{\mathcal H}_1^{T}(-k)\tau_x$, but, in addition, it is characterized by chiral symmetry, ${\mathcal H}_1(k) =-\tau_x{\mathcal H}_1(k)\tau_x$, hence the symmetry of the energy spectrum. By contrast, the effective Hamiltonian (\ref{Heff}) does not have this chiral symmetry. In essence,  this feature is due to the fact that our TI model Hamiltonian, given in the continuum limit by Eq. (\ref{HTI}), has the property ${\cal H}_{\rm TI}(k)\neq {\cal H}_{\rm TI}^T(-k)$. In addition, our numerical analysis shows that this asymmetry is significantly amplified by an oft-neglected aspect of the SC proximity effect, namely the proximity-induced renormalization of the energy spectrum, which is incorporated into the effective theory through the renormalization matrix $\widetilde{Z}$ (see Section \ref{SecIIC}). 

\begin{figure}[tbp]
\begin{center}
\includegraphics[width=0.48\textwidth]{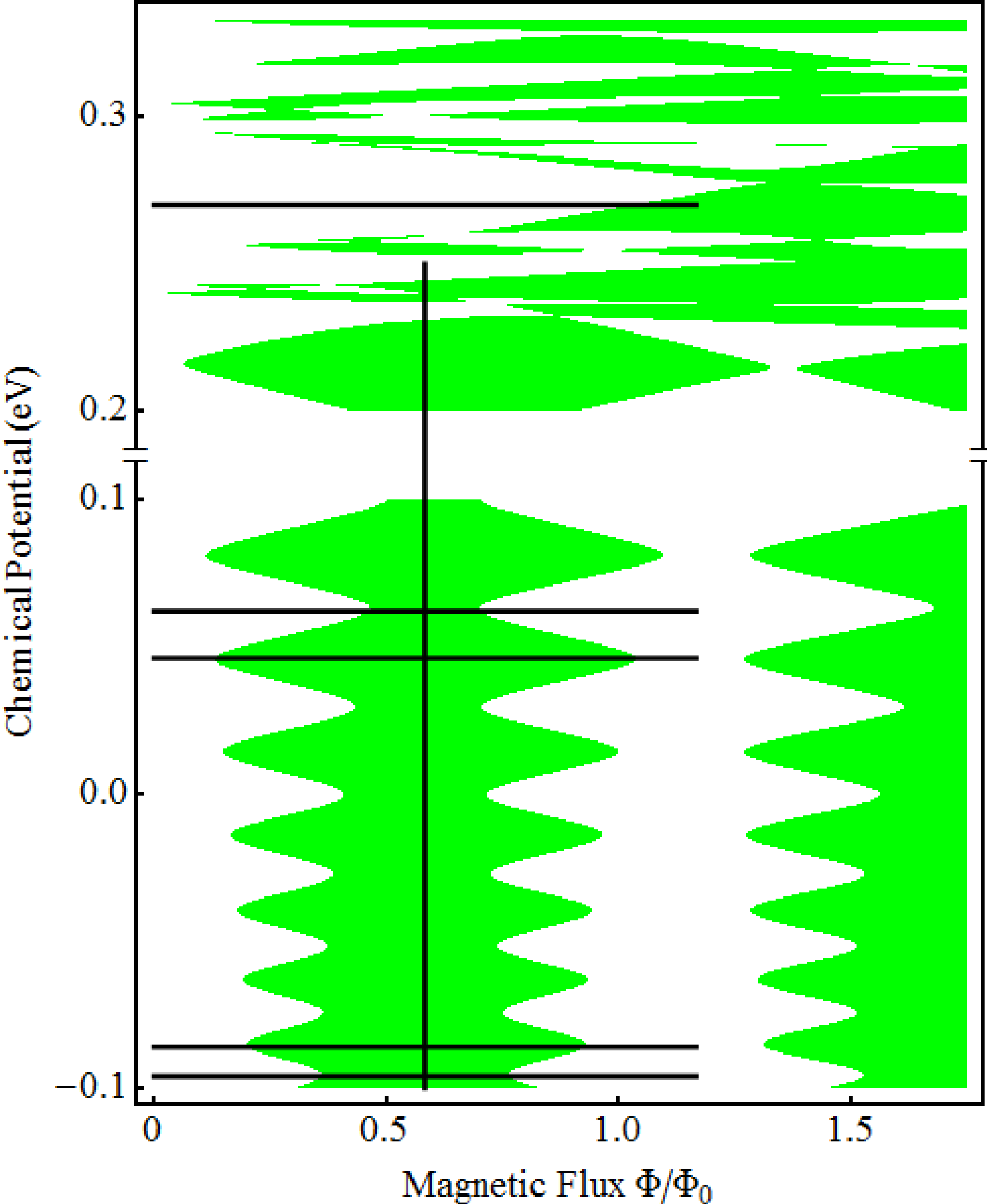}
\vspace{-8mm}
\end{center}
\caption{(Color online) Phase diagram for a SC--TI nanoribbon--SC structure (see Fig. \ref{Fig3}) as a function of the magnetic flux and the chemical potential. The green (light gray) regions correspond to a topological superconducting phase, while for parameters within the white areas the system is topologically trivial. The phase boundaries have a characteristic zigzag  shape for $\mu_{TI}$ within bulk TI gap (note that the region $0.1<\mu_{TI}<0.2$ is not shown) and breaks down when $\mu$ reaches the bulk-type bands ($\mu_{TI}>0.22$ eV). The evolution of the quasiparticle gap along the cuts corresponding to the black lines are discussed in Sec. \ref{SecIIIC}. Parameters: $L_x\times L_z =60\times 9.5$ nm, $\gamma=8\Delta_0$, $\xi=0.5$, and $\phi_{SC}=0$.}
\vspace{-3mm}
\label{Fig8}
\end{figure}

\subsection{Phase diagram}  \label{SecIIIB}

Our main interest is to determine the parameters that are consistent with the presence of a topological superconducting phase in the TI nanoribbon and to study the stability of this phase, more specifically the dependence of the induced quasiparticle gap that protects it on relevant model and control parameters. Toword that end, we map out the phase diagram of the TI-SC hybrid structure as a function of the chemical potential and applied magnetic field and determine the dependence of the phase boundaries on the size of the system, the strength $\gamma$ of the TI-SC coupling, the ratio $\xi=\tilde{t}_-/\tilde{t}_+$ of the coupling matrix elements across the interface for the two different low-energy TI orbitals, and the phase difference $\phi_{\rm{SC}}$ between bulk superconductors in structures with two interfaces (see Fig. \ref{Fig3}).  We note that, in the absence of a bias potential, structures with one and two interfaces (e.g., structures (a) and (b) in Fig. \ref{Fig3}) have similar phase diagrams and, for $\phi_{\rm{SC}}=0$, there is a    
close correspondence between a two-interface structure with coupling strength $\gamma$ and a one-interface structure with coupling strength $2\gamma$. The critical role of the bias potential will be investigated in detail in Sec. \ref{SecIIIC}. 

\begin{figure}[tbp]
\begin{center}
\includegraphics[width=0.48\textwidth]{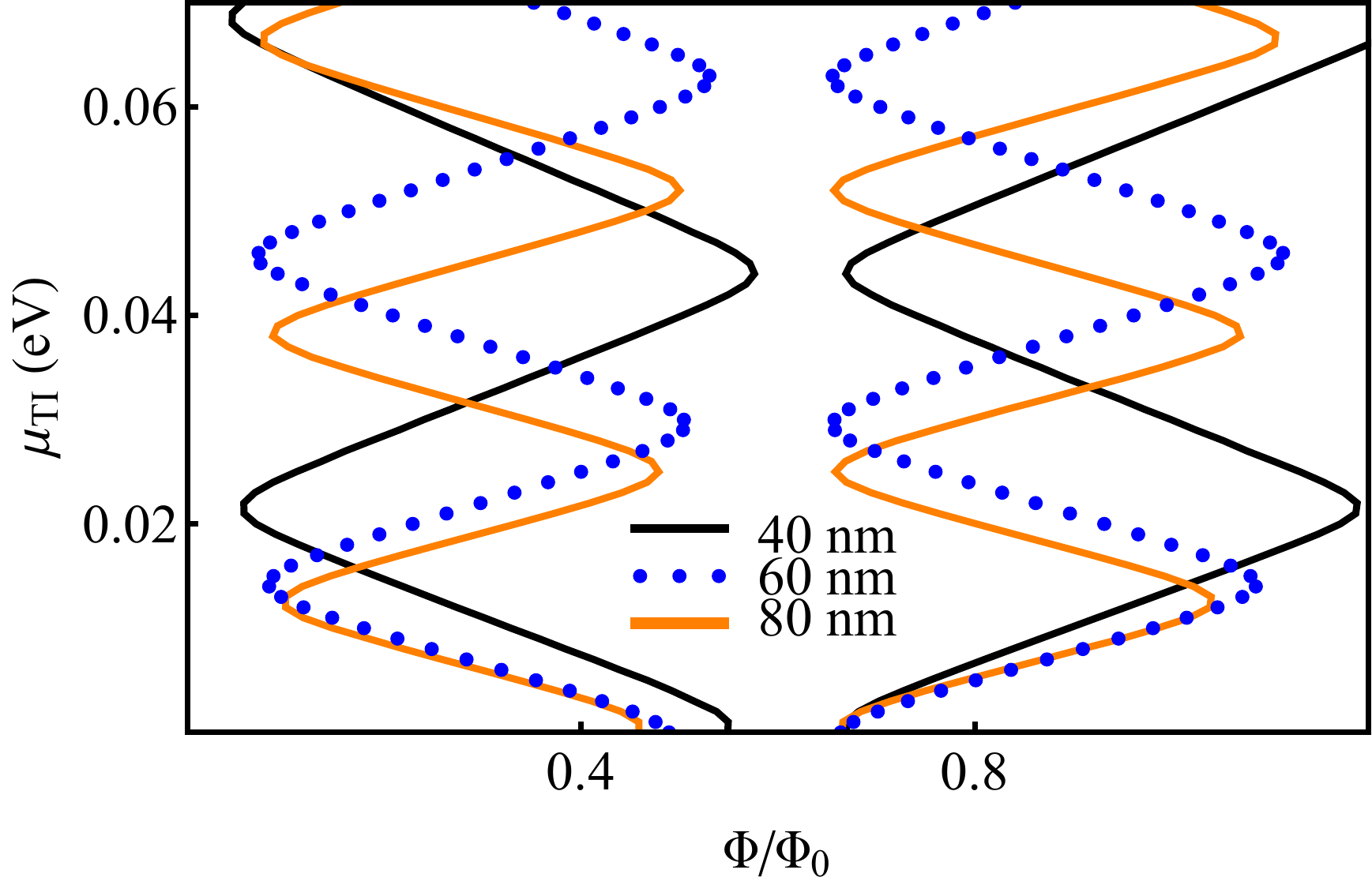}
\vspace{-8mm}
\end{center}
\caption{(Color online) Phase boundaries for proximity-coupled nanoribbons of thickness $L_x=40$ nm (black line), $L_x=60$ nm (dots), and $L_x=80$ nm (orange/light gray line). All other parameters are the same as in Fig. \ref{Fig8}. The chemical potential difference $\Delta\mu_{TI}$ between two maximum width values is controlled by the inter-band spacing. The minimum width (of the topological phase) increases with $L_x$ and the ``center line'' of the topological phase (half distance between the phase boundaries at any given value of $\mu_{TI}$) shifts toward $\Phi=0.5\Phi_0$.}
\vspace{-3mm}
\label{Fig9}
\end{figure}

A typical phase diagram is shown in Fig. \ref{Fig8}. We note the close similarity with the phase diagrams calculated in Ref. \onlinecite{Cook2012}. The phase boundaries can be obtained following Kitaev\cite{Kitaev2001} by calculating the ${\mathbb Z}_2$ topological index ${\mathcal M}$ (the Majorana number) defined as
\begin{equation}
{\mathcal M} ={\rm sgn}[{\rm Pf}B(0)]{\rm sgn}[{\rm Pf}B(\pi/\sqrt{3}a)], 
\end{equation}
where ${\rm Pf}B(k)$ is the Pfaffian of the antisymmetric matrix $B(k) = H_{\rm eff}(k) U$ defined in terms of the effective BdG Hamiltonian and the unitary operator $U$, with $\Theta = UK$ being the particle-hole symmetry anti-unitary operator and $K$ representing the complex conjugation. The Pfaffians are evaluated at the particle-hole invariant points $k=0$ and $k=\pi/\sqrt{3}a$ (the edge of the Brillouin zone for our quasi-1D system) where $H_{\rm eff}(-k)=H_{\rm eff}(k)$. The change of the Majorana number corresponds to a topological quantum phase transition between the trivial (${\mathcal M}=1$) and the nontrivial (${\mathcal M}=-1$) phases. Since ${\rm Det}\,H_{\rm eff} = {\rm Pf}\,B^2$, a sign change of the Pfaffian is accompanied by the quasiparticle gap closing at the corresponding momentum. Furthermore, in our model the gap at $k=\pi/\sqrt{3}a$ is large (of the order of 1eV) and does not close as the chemical potential and the magnetic field are varied within ranges that are relevant for the phase diagram. Consequently, the phase boundaries coincide with the closing of the quasiparticle gap at $k=0$. 

\begin{figure}[tbp]
\begin{center}
\includegraphics[width=0.48\textwidth]{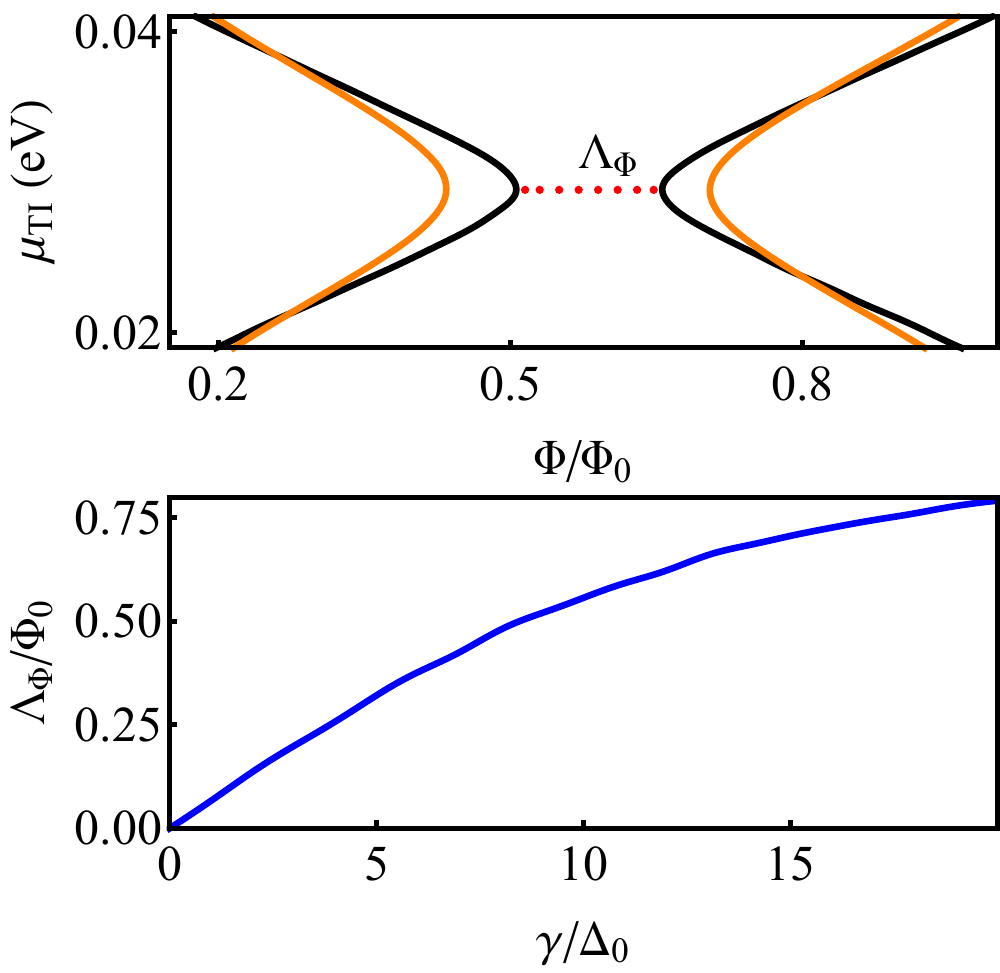}
\vspace{-8mm}
\end{center}
\caption{(Color online) {\em Top}: Phase boundaries in the vicinity of a minimum width of the topological region for $\gamma=4\Delta_0$ (black) and $\gamma=8\Delta_0$ (orange/light gray). {\em Bottom}: Dependence of the minimum width $\Lambda_{\Phi}$ on the strength of the TI-SC coupling.}
\vspace{-5mm}
\label{Fig10}
\end{figure}

The zigzag shape of the phase boundaries is related to the dependence of the $k=0$ band energies, $E_n(0)$, on the magnetic field\cite{Cook2012}, and the characteristic energy scales in the vertical direction (i.e. $\Delta\mu_{TI}$ separating two maximum width regions) are given by the inter-band gaps.  Changing the  size of the nanoribbon modifies the inter-band spacing and, consequently, the characteristic $\Delta\mu_{TI}$. This property is illustrated in Fig. \ref{Fig9}, which compares the phase diagrams corresponding to three different values of the nanoribbon width. In the regime controlled by surface-type states, the topologically trivial and non-trivial regions of the phase diagram have widths that oscillate as a function of the chemical potential (see Fig. \ref{Fig8}). The magnetic flux difference  corresponding to the minima of this width (see Fig. \ref{Fig10}, upper panel), $\Lambda_{\Phi}(\mu_{min})$, where $\mu_{min}$ takes the appropriate set of discrete values, represents a good parameter to characterize the phase diagram. Note that $\Lambda_{\Phi}(\mu_{min})$ decreases with $\mu_{min}$ and, for comparable values of the chemical potential, the values of $\Lambda_{\Phi}(\mu_{min})$ for the two phases are practically the same.  Also, as shown in Fig. \ref{Fig9},   $\Lambda_{\Phi}$ increases with $L_x$, which is another argument in favor of having thicker nanoribbons, in addition to the advantage of realizing the $\Phi=0.5\Phi_0$ conditions at lower values of the magnetic field. 

\begin{figure}[tbp]
\begin{center}
\includegraphics[width=0.48\textwidth]{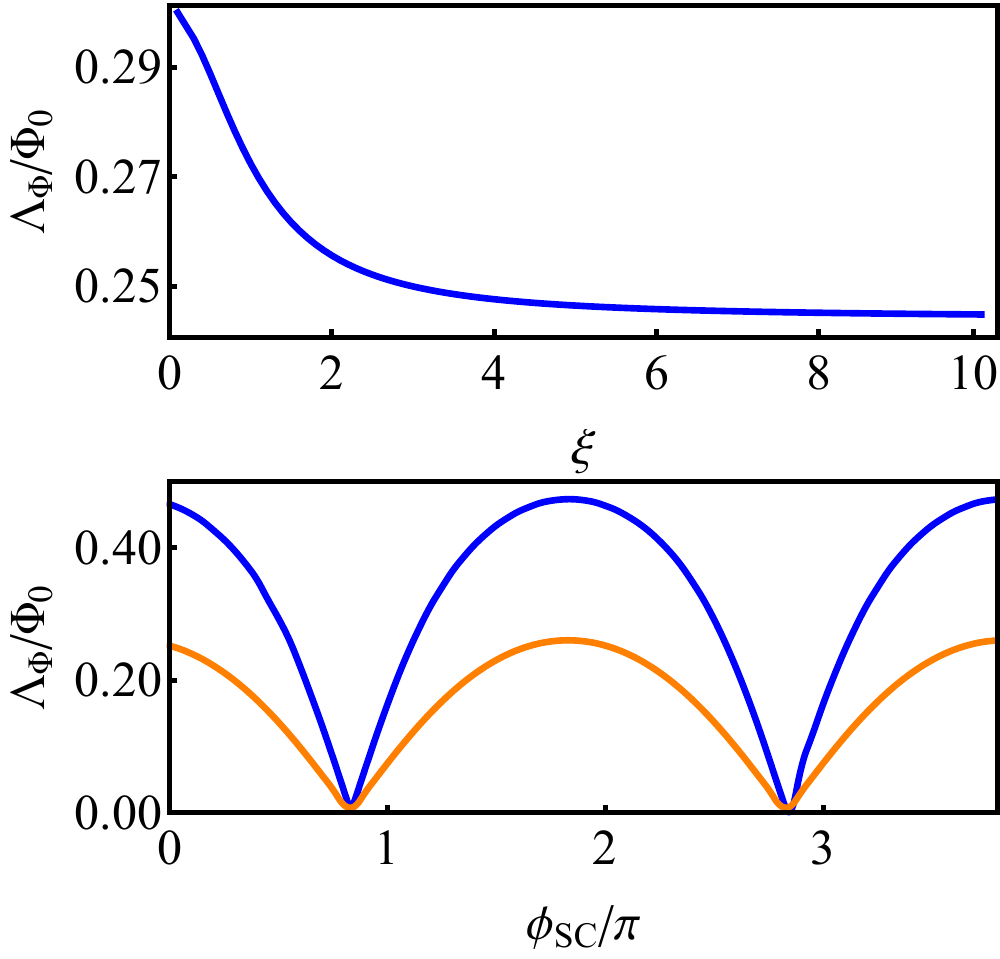}
\vspace{-8mm}
\end{center}
\caption{(Color online) {\em Top}: Dependence of the minimum width on the ratio $\xi=\tilde{t}_-/\tilde{t}_+$ of the interface hopping matrix elements for the antisymmetric and symmetric molecular orbitals (see Sec. \ref{SecIIC}). {\em Bottom}: Dependence of the minimum width on the phase difference between the bulk SCs for $\gamma=8\Delta_0$ (blue) and $\gamma=4\Delta_0$ (orange/light gray). Notice the oscillatory behavior and the nearly zero values of $\Lambda_{\Phi}$ for $\phi_{SC}\approx (0.8+2n)\pi$.}
\vspace{-5mm}
\label{Fig11}
\end{figure}

The effect of varying the strength of the TI-SC coupling on the shape of the phase boundaries can be easily summarized using the parameter  $\Lambda_{\Phi}$. As shown in Fig. \ref{Fig10}, the minimum width, which vanishes at zero coupling, as expected, increases monotonically with $\gamma$. Furthermore, we find that the phase boundaries depend weakly on the model parameter $\xi=\tilde{t}_-/\tilde{t}_+$, as shown in the top panel of Fig. \ref{Fig11}. This behavior is due to the fact that the surface-type states, which are linear combinations of states from the top (bottom) of the valence (conduction) band, contain approximately equal parts of $\lambda=+1$ and $\lambda=-1$ molecular orbitals given by Eq. (\ref{states}). More interestingly, in SC-TI-SC structures (see Fig. \ref{Fig3}) we find an oscillatory dependence of the minimum width $\Lambda_{\Phi}$ on the phase difference $\phi_{\rm{SC}}$ between the two superconductors. The results are shown in Fig. \ref{Fig11} (bottom panel). Remarkably, for phase differences  approximately equal to odd multiples of $\pi$ the minimum width of the topological region practically vanishes. This suggests the possibility of driving a topological phase transition by controlling the phase difference $\phi_{\rm{SC}}$. For example, with $\gamma=8\Delta_0$ and $\mu_{\rm{TI}}=\mu_{\rm{min}}$ one can fix the magnetic field at a value corresponding to $\Phi=0.4\Phi_0$ and drive the system from a topological SC phase ($\phi_{\rm{SC}}=0$) into a trivial phase  ($\phi_{\rm{SC}}\approx \pi$) by changing the phase difference between the bulk SCs.

\vspace{-3mm}

\subsection{Proximity-induced gap}   \label{SecIIIC}

We turn now to the main problem that we want to address in this study: the robustness of the topological superconducting phase. It was previously pointed out\cite{Potter2011} that TI-based hybrid structures can potentially have significant  advantages over semiconductor-based Majorana structures because i) there are no limitations on the magnitude of the proximity-induced quasiparticle gap in the topological phase due to (relatively) weak spin-orbit coupling strength (which is the main problem for SC-based structures), and ii) the induced SC state has enhanced immunity against time-reversal invariant disorder.  In semiconductor structures with Rashba spin-orbit coupling, for typical values of the Rashba coefficient, the quasiparticle gap in the topological phase is  significantly smaller than the induced gap at zero magnetic field.\cite{Sau2010a,Stanescu2011} This limitation is absent in TI-based systems because of the strong spin-orbit coupling that characterizes these materials. The second potential advantage stems from the fact that the low-energy surface-type states in TI wires inherit some of the properties of the topological surface states of the parent 3D material, including  their robustness against non-magnetic disorder.\cite{Cook2012} However, the actual realization of these potential advantages depends critically on the existence of a large induced gap for $B=0$, i.e. in the topologically trivial phase. If this gap is small, strong spin-orbit coupling and spin textures that minimize back scattering become irrelevant and the topological phase, together with the zero-energy Majorana bound states, will collapse in the presence of disorder and other perturbations. Previous studies assumed that the $B=0$ induced gap is a constant independent of the band index and the model parameters. However, our analysis reveals that this quantity is proportional to the wave-function amplitude at the TI-SC interface, $|\psi_n(i_0)|^2$, a quantity that is strongly dependent on the band index and, most importantly, on any bias potential acting on the nanowire. Below, we discuss the consequences of this dependence.

\begin{figure}[tbp]
\begin{center}
\includegraphics[width=0.48\textwidth]{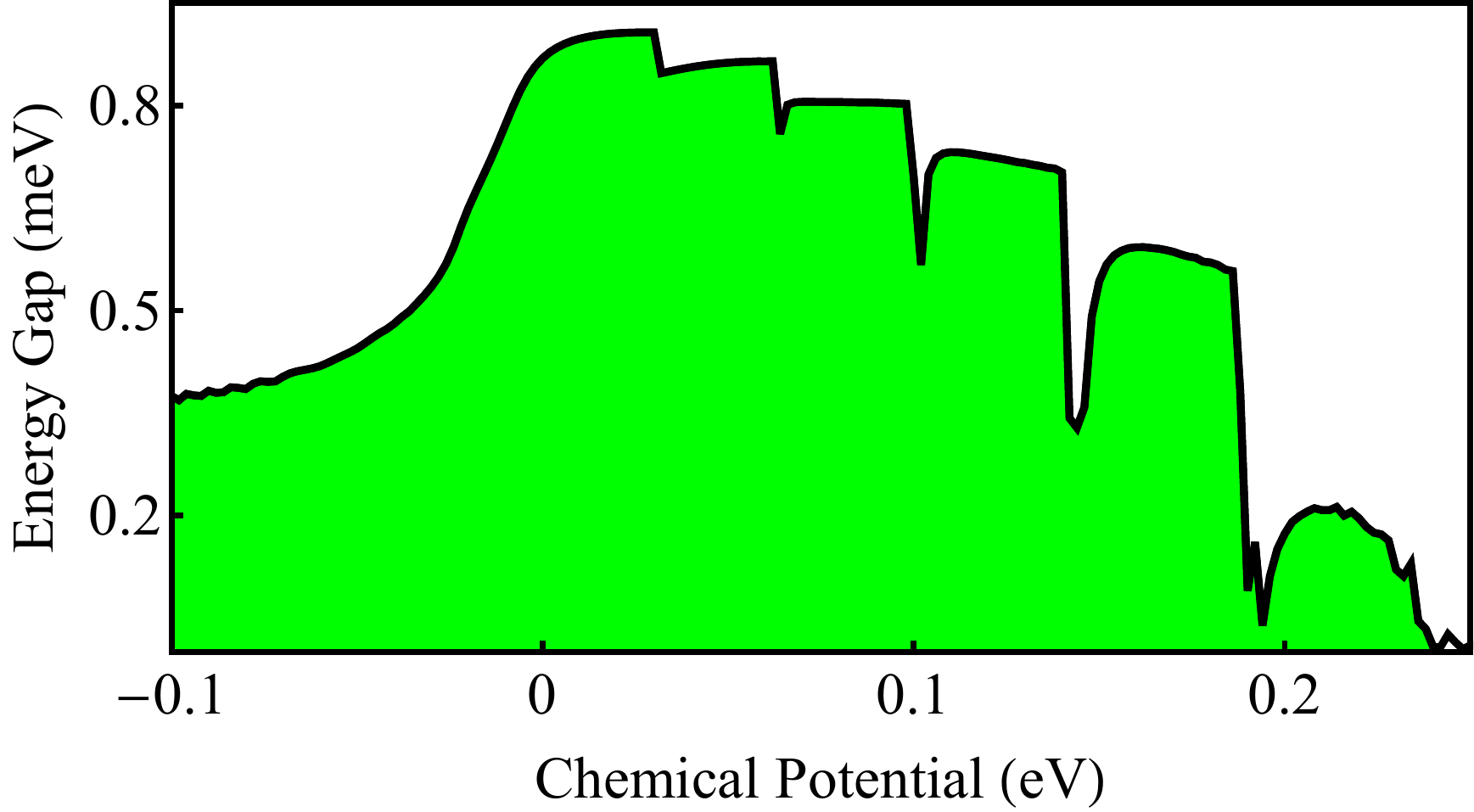}
\vspace{-8mm}
\end{center}
\caption{(Color online) Quasiparticle gap as function of the chemical potential corresponding to the vertical cut through the topological phase shown in Fig. \ref{Fig8}.  The system,  corresponding to the SC-TI-SC structure (b) in Fig. \ref{Fig3}, is characterized by the following parameters:  $L_x\times L_z =60\times 9.5$ nm, $\gamma=4\Delta_0$, $\xi=0.5$, $\phi_{SC}=0$.}
\vspace{-5mm}
\label{Fig12}
\end{figure}

We start with the ``ideal'' case of a nanoribbon sandwiched between two superconductors, as represented schematically in Fig. \ref{Fig3}, structure (b). 
In this type of structure, the bias potential should vanish by symmetry, but, even in the presence of such a bias, all surface-type states will have a large amplitude at least near one of the interfaces (see Fig. \ref{Fig5} and Fig. \ref{Fig6}). Consequently,  the effective TI-SC coupling will be strong and the corresponding induced pair potential will be large within a broad parameter range. In Fig. \ref{Fig12}, we show the dependence of the induced quasiparticle gap on the chemical potential corresponding to a vertical cut though the topological phase, as shown in Fig. \ref{Fig8}. There are several important features that we want to emphasize. First, if we consider as our reference the value of the induced superconducting gap (at zero magnetic field) currently achievable in semiconductor-based structures,\cite{Mourik2012} i.e. approximately $0.25$ eV, we note that the quasiparticle gap in the topological phase of this TI-SC hybrid system is large for, practically,  any value of the chemical potential within the bulk TI gap. The collapse of the gap for $\mu_{TI} >0.24$ eV is due to the fact that the bulk-type bands start to become occupied. Second, the smaller gap at negative energies is associated with the top-most negative-energy band, which is nearly doubly degenerate away from  $k=0$ (see Fig. \ref{Fig4}). Since in real systems the Dirac point is very close to the top of the valence band (a feature that is not quantitatively captured by the model used in these calculations because it misses some finite $k$-vector states at the top of the valence band\cite{Zhang2009}), the regime $\mu_{\rm{TI}}<0$ does not have any significant practical importance. 
Third, the sharp drops correspond to the chemical potential reaching the bottom of a pair of bands that are not exactly degenerate.  This is a finite-size effect that characterizes thin nanoribbons and is due to the fact that the $k=0$ degeneracy condition is not realized at the same value of the magnetic field for all bands (see Fig. \ref{Fig4}). Setting the magnetic field to a value that corresponds to the exact degeneracy of the lowest-energy bands will leave small gaps at $k=0$ between the higher-energy bands. By increasing the size of the system the degeneracy condition approaches  the band-independent expression $\Phi=0.5 \Phi_0$, and the sharp minima of the quasiparticle gap disappear. 

\begin{figure}[tbp]
\begin{center}
\includegraphics[width=0.48\textwidth]{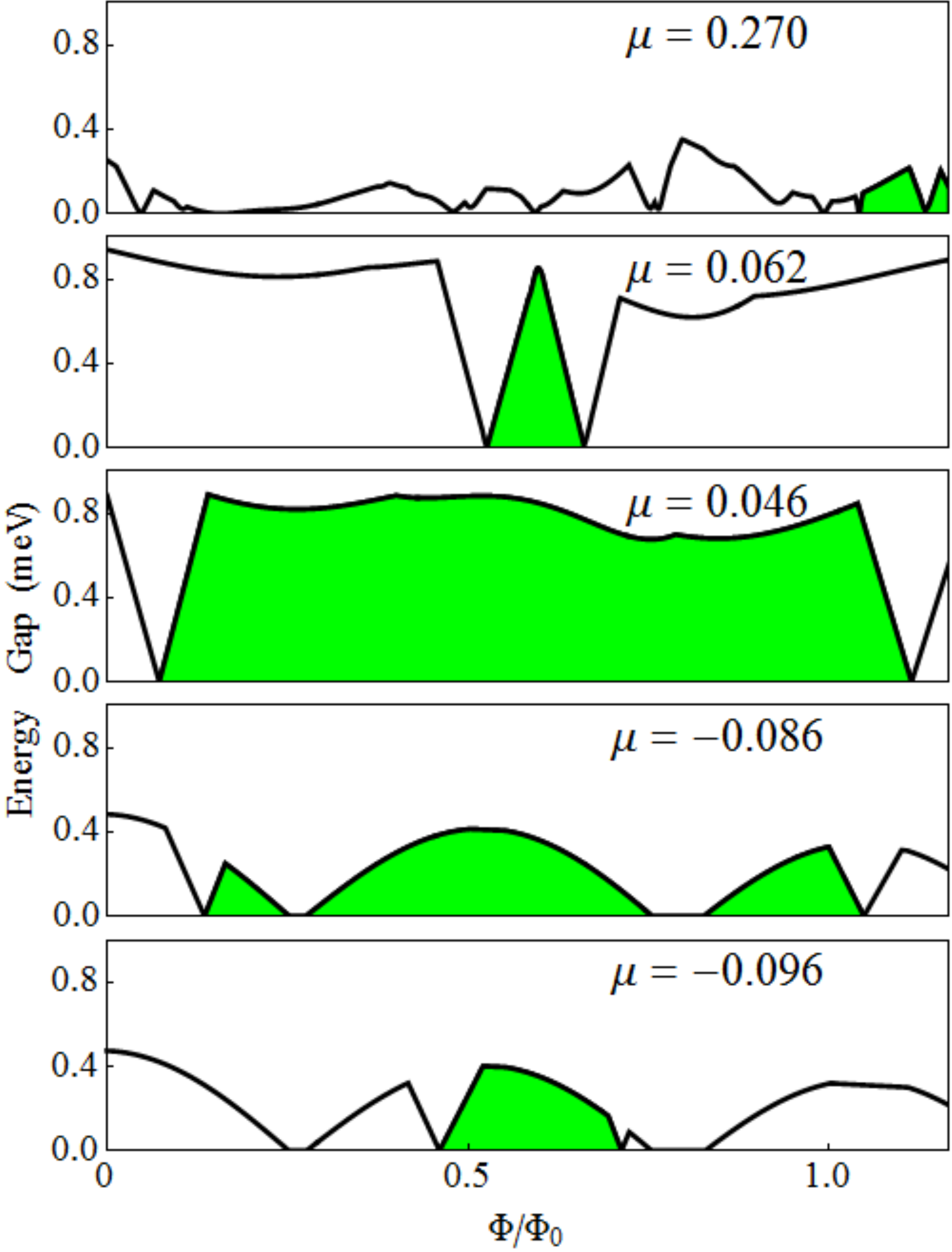}
\vspace{-8mm}
\end{center}
\caption{(Color online) Quasiparticle gap as a function of the magnetic flux for different values of the chemical potential corresponding to the horizontal cuts in Fig. \ref{Fig8}. The model parameters are the same as in Fig. \ref{Fig12}. The shaded regions correspond to the topological superconducting phase. The quasiparticle gap has a characteristic V-type shape in the vicinity of a topological quantum phase transition and vanishes at the transition. Note that for some values of the chemical potential (e.g., $\mu=-0.086$ eV), the vanishing of the gap is not associated with a phase transition.}
\vspace{-5mm}
\label{Fig13}
\end{figure}

To gain a better understanding of the low-energy properties characterizing different regions of the phase diagram, we also calculate the dependence of the quasiparticle gap on the magnetic field for different values of the chemical potential corresponding to the horizontal cuts in Fig. \ref{Fig8}. The results are shown in Fig. \ref{Fig13}. We note that the vanishing of the gap associated with a V-shaped dependence on the magnetic flux signals a topological quantum phase transition between the topologically trivial and nontrivial phases. In the vicinity of a transition, the lowest-energy state is always the $k=0$ state from the top occupied band. In fact, the characteristic V-shape results from the nearly linear dispersion of the $k=0$ states with the magnetic flux.  Interestingly, for some values of the chemical potential it is possible to have zero quasiparticle gap over a finite magnetic-field range (see Fig. \ref{Fig13}, lowest two panels). We emphasize that in these cases there is no associated topological quantum phase transition, since the gapless states have finite $k$-vectors (see Fig. \ref{Fig7}, lower panel, and the discussion in Sec. \ref{SecIIIA2}).  The corresponding gapless superconducting phases can be either topologically trivial (e.g., $\mu=-0.096$ eV in Fig. \ref{Fig13}) or nontrivial  ($\mu=-0.086$ eV). Finally, we note the sharp maximum that characterizes the $\mu=0.062$ eV cut though a narrow section of the topological region. The position of the peak is determined by the $k=0$ degeneracy condition ($\Phi\gtrsim 0.5\Phi_0$), which for narrow wires is slightly band-dependent. Hence, when considering a constant field cut, as in Fig. \ref{Fig12}, the maxima will not be perfectly aligned and the quasiparticle gap will have a sharp drop whenever   the chemical potential passes through a narrow section of the topological region. Furthermore, since the width of a narrow section of the topological region depends strongly on the TI-SC coupling (see Fig. \ref{Fig10}),  for $\gamma$ less than a certain critical value, $\gamma_c\approx 4\Delta_0$, the quasiparticle gap in this region is determined by the energy of the $k=0$ state from the top occupied band (which results in a quasilinear dispersion and a sharp peak, as in Fig. \ref{Fig13}), while for $\gamma>\gamma_c$ the quasiparticle gap is determined by a $k\neq 0$ state, which disperses weakly with the magnetic field and generates a peak with a nearly flat top. Consequently, the sharp drops in the quasiparticle gap as a function of the chemical potential (like those in Fig. \ref{Fig12}) can be significantly attenuated, even in narrow ribbons, by increasing the TI-SC coupling constant, since the perfect alignment of nearly flat top peaks is not a problem. On the contrary, in weakly coupled structures these features are very pronounced and could be a problem as, in the presence of disorder,  they become a source of low-energy states.     

\begin{figure}[tbp]
\begin{center}
\includegraphics[width=0.48\textwidth]{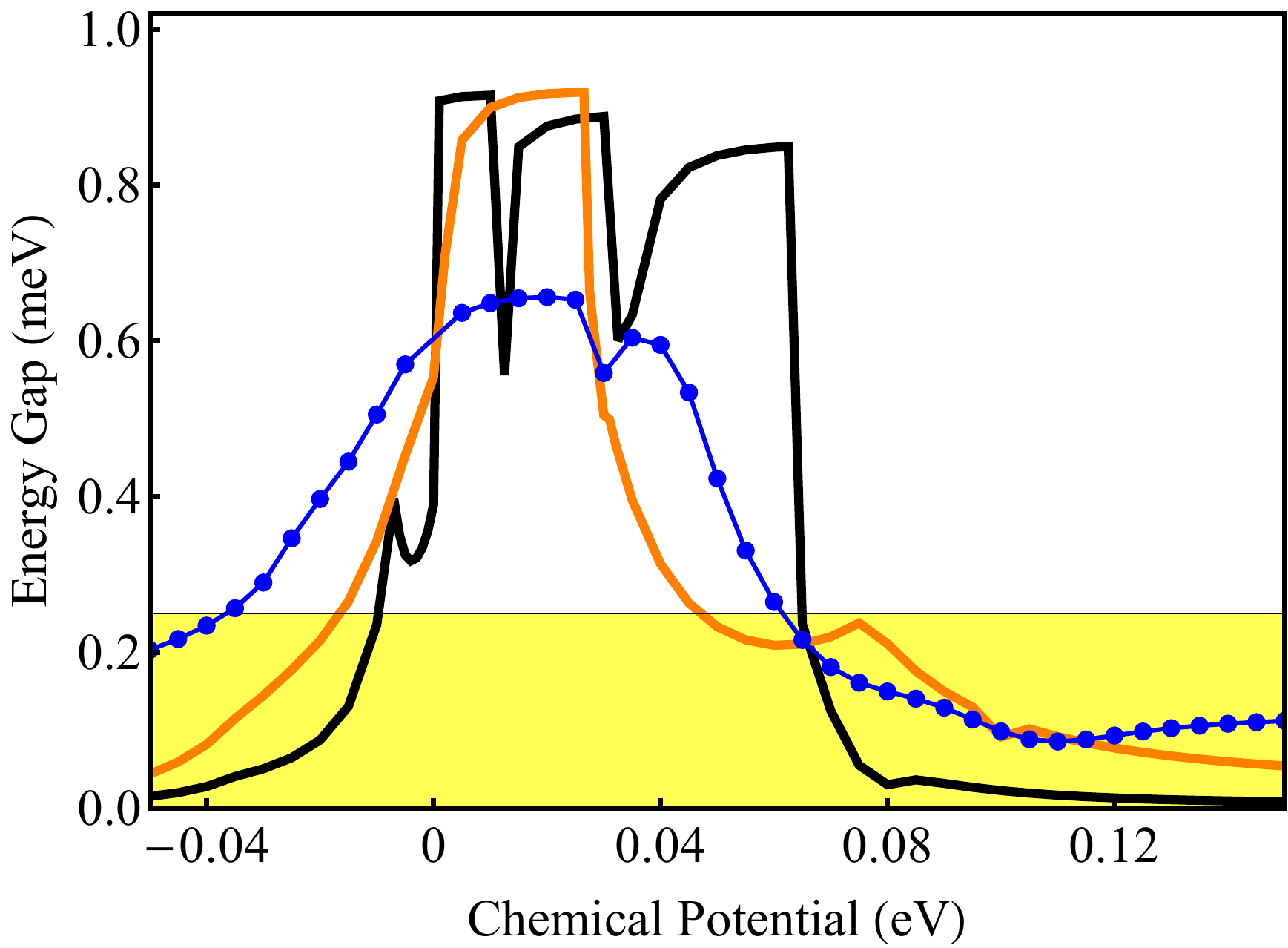}
\vspace{-8mm}
\end{center}
\caption{(Color online) Quasiparticle gap as a function of the chemical potential for a TI-SC system with one interface-structure (a) in Fig. \ref{Fig3}--and bias potential $V_{\rm max}=0$ (thin line and dots),  $V_{\rm max}=0.03$ eV [orange (light gray) line], and $V_{\rm max}=0.06$ eV (black). Unlike the case of a symmetric SC-TI-SC structure (see Fig \ref{Fig12}), the range of chemical potential corresponding to an induced gap larger than the ``reference`` value ($0.25$ meV, shaded area) is rather small and corresponds to low band occupancy (single-band occupancy for $V_{\rm max}=0.03$ eV). Outside this range, the gap practically collapses with increasing bias potential.}
\vspace{-5mm}
\label{Fig14}
\end{figure}

Next, we turn our attention toward hybrid structures with a single TI-SC interface and focus on the effect of a bias potential.  Specifically, we consider a TI-SC system such as structure (a) in Fig. \ref{Fig3}. This type of structure will intrinsically have a proximity-induced bias potential arising from the $\zeta$ term in Eq. (\ref{Gsc}).  An additional contribution may be present at the interface with the substrate (not considered explicitly). As discussed in Sec. \ref{SecIIC}, we include these interface contributions to the bias potential in  the $H_V$ term of the Hamiltonian and study their effect by controlling the independent bias parameter $V_{\rm max}=V(N_z)-V(1)$ in Eq. (\ref{Vi}). We emphasize again that the details of the dependence on the quintuple layer index $i_z$ in Eq. (\ref{Vi}) are not important, because surface-type states have most of their weight on the top and bottom layers. 
We consider three different cases, one corresponding to zero bias potential,  $V_{\rm max}=0$,  and two corresponding to  $V_{\rm max}=0.03$ eV and 
 $V_{\rm max}=0.06$ eV, respectively. The dependence of the quasiparticle gap on the chemical potential for these three cases is shown in Fig. \ref{Fig14}. When comparing these results with those shown in Fig. \ref{Fig12}, the first striking feature is the drastic reduction of the chemical potential range corresponding to values of the quasiparticle gap larger than the ``reference'' value of $0.25$ meV. In fact, this range is limited by a low band occupancy condition, more precisely the requirement of having  one pair of Fermi points for $V_{\rm max}=0.03$ eV and one or three pairs for $V_{\rm max}=0.06$ eV.  The physics responsible for this behavior in the presence of a finite bias can be easily understood in terms of the real-space structure of the low-energy states (see Fig. \ref{Fig6}). More specifically, the bias potential determines the low-energy surface-type states to have significant amplitude near either the top or the bottom surface of the nanoribbon. Consequently, any band containing states with a small amplitude at the interface will be very weakly coupled to the superconductor and will be characterized by a small induced gap. By contrast, bands containing states with  large amplitude at the interface will be characterized by large induced gaps. The top negative energy band (see Fig. \ref{Fig4}) is a weakly coupled band and this explains the sharp drop of the induced gap for $\mu_{\rm{TI}}<0$. Similarly, the collapse at positive values of the chemical potential is due to the fact that the first weakly coupled positive-energy band becomes occupied. This is the third positive-energy band for $V_{\rm max}=0.03 $ eV and the fifth for $V_{\rm max}=0.06$ eV. 	 

Understanding the rapid decrease of the induced gap for values of the chemical potential that do not satisfy the single-band occupancy condition in the unbiased system, $V_{\rm max}=0$, is less obvious and is due to a more subtle mechanism. Since the surface-type states are now equally distributed on the top and bottom surfaces (see Fig. \ref{Fig5}), one would expect a significant gap over a larger chemical potential range. However, away from $\mu=0$ both the top negative-energy band and the lowest positive-energy band are nearly doubly degenerate (see Fig. \ref{Fig4}). When the nanoribbon is proximity-coupled to the SC, the nearly degenerate bands combine into a strongly coupled mode and a weakly coupled mode, which is responsible for the reduced quasiparticle gap. We emphasize that this mechanism does not involve any proximity-induced bias contribution. In other words, the normal term  proportional to $\zeta$ from Eq. (\ref{Gsc}) is completely neglected. 

\begin{figure}[tbp]
\begin{center}
\includegraphics[width=0.47\textwidth]{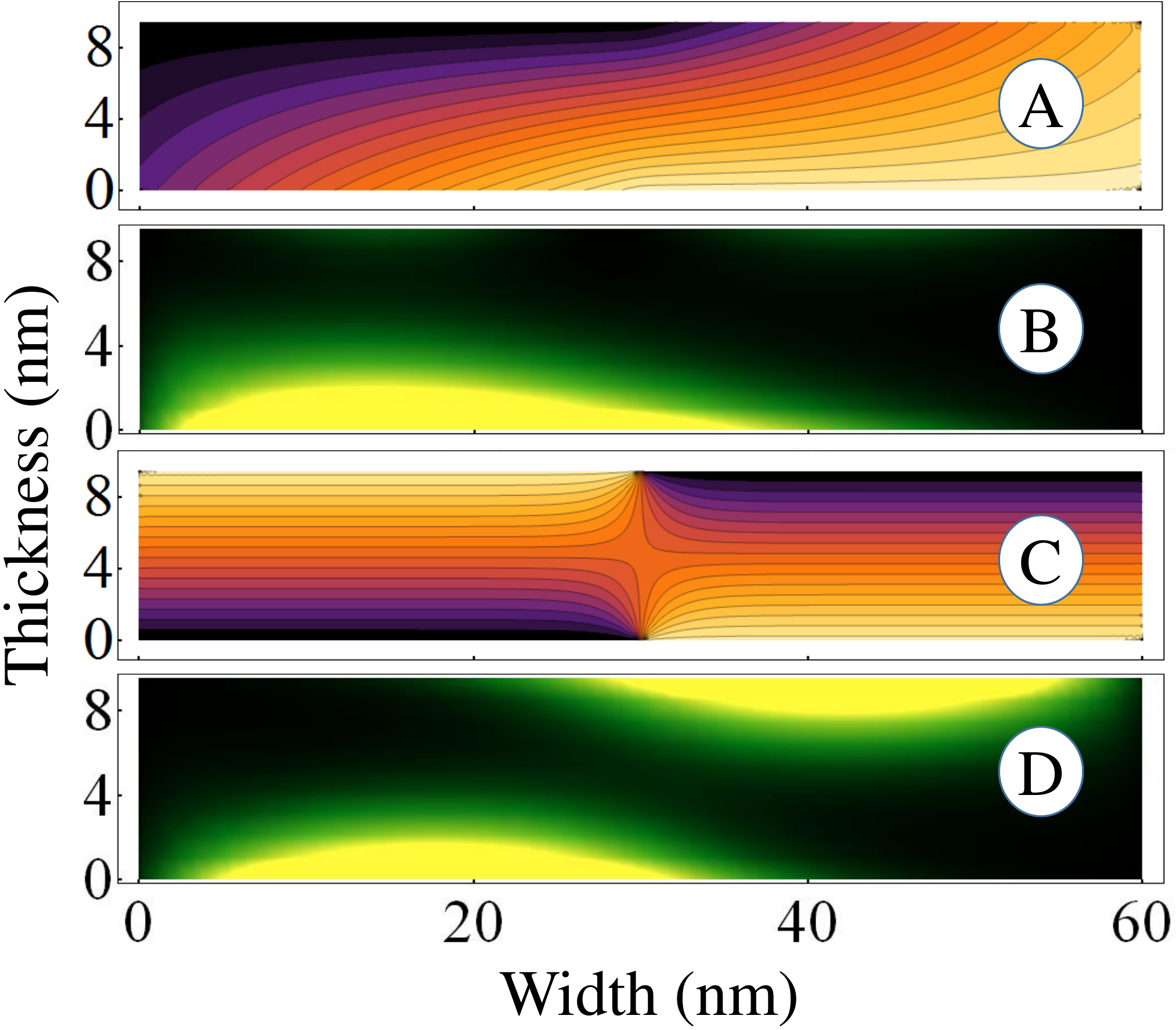}
\vspace{-8mm}
\end{center}
\caption{(Color online) Cross section profile of the applied gate potential corresponding to hybrid structure (c) from Fig. \ref{Fig3} (panel A) and structure (d) (panel C). The yellow (light gray) color represents the gate potential $V_{\rm gate}$, while the dark region corresponds to the value of the potential at the TI-SC interface, $V(i_0)=0$. Panels B and D show the real space structure of some representative low-energy states supported by the TI beribbon in the presence of the gate potential. For the single-gate structure (panel A), some states are localized near the TI-SC interface, others (see panel B) away from the interface. For the two-gate structure (panel C) all states have some weight near the two TI-SC interfaces  (panel D).}
\vspace{-5mm}
\label{Fig15}
\end{figure}

\begin{figure}[tbp]
\begin{center}
\includegraphics[width=0.48\textwidth]{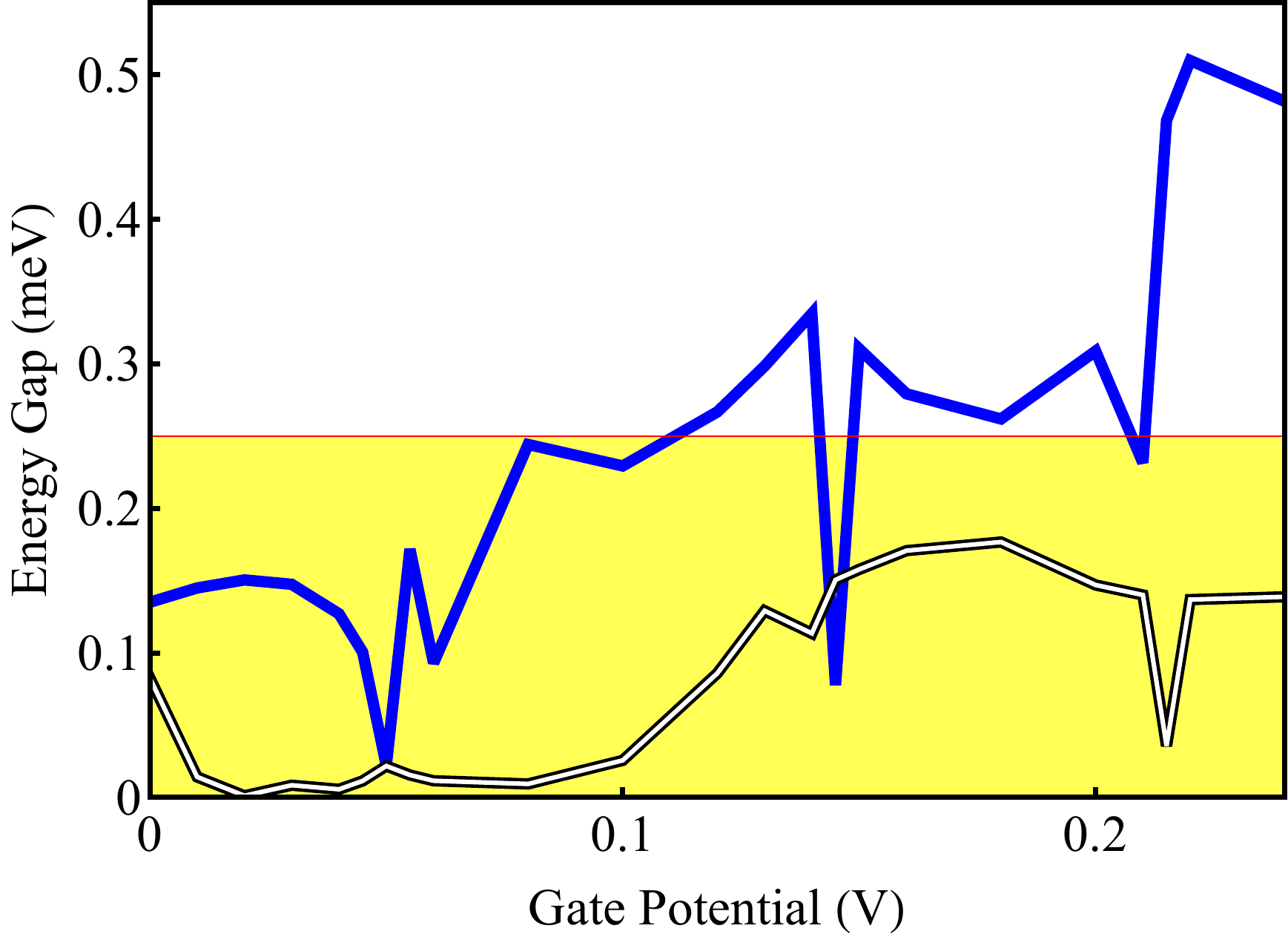}
\vspace{-8mm}
\end{center}
\caption{(Color online) Dependence of the induced quasiparticle gap on the applied gate potential for a single-gate structure with a potential profile like that in Fig. \ref{Fig15}, panel A (double thin line), and a double-gate structure with potential profile given in  Fig. \ref{Fig15}, panel C [full blue (dark gray) line]. The chemical potential is fixed at the value $\mu_{\rm{TI}}=0.22$ eV, close to the bulk TI gap edge. Applying a gate potential shifts the spectrum up in energy, but it also distorts it. For the single-gate structure, at least one weakly coupled band remains occupied and limits the size of the induced gap.}
\vspace{-5mm}
\label{Fig16}
\end{figure}

The final aspect that we want to address concerns the possibility of controlling the chemical potential using applied gate potentials. Although for a symmetric SC-TI-SC structure there are no strict requirements concerning the chemical potential (see Fig. \ref{Fig12}), as long as it is not too close to the bulk TI gap edge, this is not the case for single-interface structures, as illustrated by the results shown in Fig. \ref{Fig14}. For concreteness, we consider two cases corresponding to structures (c) and (d) in Fig. \ref{Fig3}. We fix the chemical potential to a value below to the bulk TI gap edge, $\mu_{\rm{TI}}=0.22$ eV, and we determine the profile of the applied potential by assuming that its value in the vicinity of the TI-SC interface will be independent of the gate voltage due to strong screening by the superconductor. The profiles for the two structures are shown in Fig. \ref{Fig15}, panels A and C, respectively. The dependence of the induced gap on the gate potential can be anticipated by analyzing the evolution of the spectrum and the real-space properties of the low-energy wave functions. We find that for the single-gate structure, there is at least one weakly-coupled band that remains occupied. The transverse profile for a representative state that belongs to a weakly coupled mode is shown shown in Fig. \ref{Fig15}, panel B. Note that the TI-SC interface in this structure corresponds to the left half of the top surface, a region where the amplitude of the state is extremely small. By contrast, in the case of the symmetric two-gate structure, we find that, for a strong-enough gate potential, all occupied bands are strongly-coupled, as the corresponding states have significant weight near the TI-SC interfaces (see Fig. \ref{Fig15}, panel D). Explicit calculation of the dependence of the quasiparticle gap on the gate potential confirms these expectations. As shown in Fig. \ref{Fig16}, the induced gap does not exceed the reference value ($0.25$ meV) in the single-gate structure, but it can be made large using the symmetric structure by applying a strong enough gate potential.   This strengthens our previous conclusions concerning the limitations of single-interface TI-SC structures. In essence, we can summarize the key steps that led us to uncovering these limitations as follows. i) The SC proximity effect can be incorporated into an effective low-energy theory through interface contributions to the BdG Hamiltonian. ii) The effective TI-SC coupling is determined by the amplitude of the low-energy wave function at the interface and, consequently, is strongly band-dependent. iii) The low-energy states corresponding to energies within the bulk TI gap have surface-type character, with most of their weight near either both or only one of the top and bottom surfaces of the nanoribbon. iv) Single-interface TI-SC hybrid structures have a rather limited potential to host robust topological SC phases and Majorana bound states because, for arbitrary values of the chemical potential within the bulk TI gap, one or more weakly-coupled bands 
become occupied. These weakly-coupled bands represent a source of low-energy excitations that can compromise the robustness of the Majorana zero-energy modes. We find that  realizing two-interface structures, e.g., by sandwiching the TI nanoribbon between two SCs, can cure this problem.  

\vspace{-3mm}

\section{Conclusions} \label{SecIV}

\vspace{-1mm}

In this work we study the low-energy spectrum of topological insulator (TI) nanoribbons proximity-coupled to $s$-wave superconductors (SCs) using a four-band tight-binding model to describe the TI ribbon and explicitly incorporating the proximity effects induced by the bulk SC. We demonstrate that the low-energy properties of the hybrid system can be described using an effective Bogoliubov -- de Gennes Hamiltonian that incorporates the superconducting proximity effect through i) an anomalous interface contribution responsible for the induced pair potential, ii) a normal interface term that generates a proximity-induced bias potential, and iii) an energy-renormalization matrix. We show that the strength of the superconducting proximity effect is determined by an effective TI-SC coupling that depends on the transparency of the interface, which is parametrized in  our theory by two interface hopping matrix elements, $\tilde{t}_+$ and $\tilde{t}_-$, and on the amplitude of the low-energy quantum states of the TI nanoribbon at the interface. We investigate the real-space properties of these low-energy states and show that the states with energies within the bulk TI gap have surface-type character, i.e., large amplitude near the boundaries of the TI nanoribbon and exponentially vanishing amplitude toward the core of the wire. We find that, in the presence of a bias potential, these surface-type states have most of their weight near the top or bottom surface of the nanoribbon. Consequently, the superconducting proximity effect, including the induced pair potential, will be enhanced for low-energy bands with states localized at the TI-SC interface and will be strongly suppressed for bands with states localized near the opposite surface. For nearly degenerate bands, the splitting into a strongly paired and a weakly paired mode occurs even in the absence of a bias term in the Hamiltonian. 

 We investigate the potential impact of various mechanisms that control the strength of the superconducting proximity effect in quasi-1D TI systems by calculating the dependence of the topological phase diagram and the induced quasiparticle gap on relevant model and control parameters, including the chemical potential, applied magnetic fields,  and gate potentials.  The interest in TI nanoribbon-superconductor hybrid structures  stems from the expectation that they have the ability to harbor exceptionally robust topological superconducting phases and zero-energy  Majorana bound states. This robustness rests, ultimately, on the possibility of inducing large quasiparticle gaps that, for appropriate values of the magnetic flux, remain large within a wide range of values for the chemical potential. Our detailed analysis shows that this may not be a straightforward task. In particular, we find that single-interface TI-SC structures are particularly susceptible to experience the collapse of the induced gap whenever the chemical potential is far enough from the value corresponding to the bulk TI Dirac point.  Moreover, if the chemical potential is close to the bulk gap edge, changing it by using gate potentials can be rather ineffective in single interface structures, as the generated bias produces bands with sates localized away from the TI-SC interface, which results in small quasiparticle gaps. On the other hand, symmetric structures, such as a TI nanowire sandwiched between two superconductors, are capable of realizing the full potential of TI-based Majorana wires but may pose additional engineering challenges.  
 

\subsection*{Acknowledgments} 

This work was supported by WV HEPC/dsr.12.29.

\bibliography{REFERENCES}
\end{document}